\documentclass[aps,prb,amsmath,amssymb,footinbib,showpacs,twocolumn]{revtex4-2}
\usepackage{amsmath}
\usepackage{amssymb}
\usepackage{amsthm}
\usepackage{setspace}
\usepackage{graphicx}
\usepackage{braket}
\usepackage{mathrsfs}
\usepackage{float}
\usepackage{booktabs}
\usepackage[colorlinks = true,linkcolor = blue,urlcolor  = blue,citecolor = blue,anchorcolor = blue]{hyperref}
\usepackage[utf8]{inputenc}
\usepackage[english]{babel}
\usepackage{bm}
\usepackage{comment}

\begin{document}

\title{Localization in two-dimensional fermions with arbitrary pseudospin}

\author{Adesh Singh}
\affiliation{School of Physical Sciences, Indian Institute of Technology Mandi, Mandi 175005, India}
\author{Gargee Sharma}
\affiliation{School of Physical Sciences, Indian Institute of Technology Mandi, Mandi 175005, India}

\begin{abstract}
In condensed matter, limited symmetry constraints allow free fermionic excitations to exist beyond the conventional Weyl and Dirac electrons of high-energy physics. These excitations carry a higher pseudospin, naturally generalizing the Weyl fermion. How do electrons beyond the conventional Dirac and Weyl fermions localize under disorder? In this Letter, we solve the problem of localization of \textcolor{black}{two-dimensional} free fermionic excitations carrying an arbitrary pseudospin-$s$. 
We derive exact analytical expressions for fermionic wavefunctions and exploit their curious mathematical connection to Pascal's triangle to evaluate relevant quantities such as scattering time, renormalized velocity, Cooperon, and magnetoconductivity.
We discover that the gapless Cooperon mode solely depends on the pseudospin even when the Fermi surface is composed of multiple pockets, leading to weak localization (antilocalization) behavior for integer (half-integer) $s$, irrespective of the band index. Remarkably, \textcolor{black}{the localization corrections increase with $s$, but the relative localization corrections are found to decrease} with $s$, i.e., faster-moving relativistic electrons are \textcolor{black}{less} susceptible to disorder effects. \textcolor{black}{Coupled with our elementary analysis on electron-electron interactions, this sheds insights on} Anderson and many-body localization in these materials.

\end{abstract}

\maketitle
\section{Introduction}
Electrons in a periodic potential can lead to free-fermionic excitations that display striking quantum mechanical properties. A foremost example is graphene~\cite{neto2009electronic,sarma2011electronic}, where the additional sublattice degree of freedom provided by the honeycomb lattice maps its low-energy theory to that of a relativistic spin $s=1/2$ massless Dirac electron. %In graphene, `spin', actually represents a `pseudospin-1/2' since the degree of freedom is provided by the sublattice and not the physical spin.
Since the discovery of graphene, advances in material science have made it possible to realize a wide variety of fermionic excitations in systems such as topological insulators~\cite{hasan2010colloquium,qi2011topological,vafek2014dirac}, Van der Waal heterostructures~\cite{geim2013van}, Weyl and Dirac semimetals~\cite{armitage2018weyl}, topological superconductors~\cite{kitaev2001unpaired,lutchyn2010majorana,oreg2010helical}, and the much-celebrated moir\'{e} heterostructures~\cite{cao_unconventional_2018,cao_correlated_2018, yankowitz_2019_tuning,gibney2019magic,sharma2021carrier}. These can display a wide variety of fascinating electronic properties, such as mimicking the high-energy Weyl, Dirac, and Majorana fermions~\cite{vafek2014dirac,geim2013van,armitage2018weyl,kitaev2001unpaired,lutchyn2010majorana,oreg2010helical}, hosting flat bands that can facilitate correlated physics~\cite{cao_unconventional_2018,cao_correlated_2018, yankowitz_2019_tuning,gibney2019magic,sharma2021carrier}, exhibiting higher pseudospins~\cite{bradlyn2016beyond}, to name a few. The prospect of realizing these features in cold atomic lattices is a contemporary research theme~\cite{goldman2016topological,zhang2018topological}.

In high-energy physics, the constraints imposed by Poincar\'{e} symmetry make it impossible to realize fermions beyond $s=1/2$, but in periodic condensed matter systems, the constraints are lesser. Bradyln et al.~\cite{bradlyn2016beyond} realized the possibility of finding free fermionic topological excitations in condensed matter systems that have no analogs in high-energy physics. These excitations, which are stabilized by certain symmetries, carry higher-pseudospins ($s>1/2$), are $n-$fold degenerate ($n>2$), and carry a nontrivial Chern number $|\mathcal{C}|>1$~\cite{bradley1972c,wieder2016double,bradlyn2016beyond,raoux2014dia,ezawa2016pseudospin}. Furthermore, $\mathbf{k}\cdot \mathbf{p}$ theory and a corresponding low-energy $\mathbf{k}\cdot\mathbf{S}$ Hamiltonian exists for systems belonging to certain spacegroups~\cite{bradley1972c,bradlyn2016beyond} \textcolor{black}{Proposed compounds where such a low-energy $\mathbf{k}\cdot\mathbf{S}$ Hamiltonian include Ag$_3$Se$_2$Au, Pd$_3$Bi$_2$S$_2$,  MgPt, Li$_2$Pd$_3$B, CuBi$_2$O$_4$ among several others.}

Deviation from periodicity due to disorder is experimentally inevitable. Although disorder is typically not desirable, it can lead to intriguing phenomena of solely quantum origin. In the presence of strong disorder, electrons can localize leading to an Anderson insulating phase~\cite{anderson1958absence}.
Constructive wave interference in even weakly disordered solids leads to negative quantum correction to the Drude conductivity, known as weak localization (WL)~\cite{lee1985disordered,akkermans2007mesoscopic,altshuler1980magnetoresistance,bergmann1984weak,chakravarty1986weak}, which is a precursor to Anderson localization~\cite{filoche2012universal}. 

Interestingly in graphene, the pseudospin generates a Berry phase that leads to a destructive wave interference, resulting in a positive quantum correction to the conductivity~\cite{hikami1980spin,suzuura2002crossover,khveshchenko2006electron,mccann2006weak}. This phenomenon, known as weak antilocalization (WAL), was originally proposed to occur in a spin-orbit coupled two-dimensional electron gas ~\cite{hikami1980spin}, where the rotation of the physical spin causes the phase difference. Although graphene and spin-orbit coupled systems belong to the same symplectic symmetry class, their scaling behavior is remarkably different~\cite{markovs2006critical,ostrovsky2007quantum,bardarson2007one,nomura2007topological}.
Despite intensive studies on the localization of Dirac and Weyl fermions~\cite{suzuura2002crossover,khveshchenko2006electron,mccann2006weak,gorbachev2007weak,wu2007weak,gorbachev2007weak,tikhonenko2008weak,tikhonenko2008weak,tkachov2011weak,singh2023quantum,lu2011competition,lu2013intervalley,lu2014finite,fu2019quantum,markovs2006critical,ostrovsky2007quantum,bardarson2007one,nomura2007topological}, the fate of free fermionic excitations beyond the Dirac and Weyl cases under disorder remains a broadly open problem.

%Higher pseudospins emerge naturally in several other systems of interest, such as the $\alpha-\mathcal{T}_3$ lattice model~\cite{raoux2014dia}. This model comprises of a hexagonal lattice with atoms situated at the vertices of the hexagons and their centers, and thus describes a three-band system of pseudospin-1 fermions. Two-dimensional sectors of Kane fermions in Hg$_{1-x}$Cd$_x$Te and Cd$_3$As$_2$ have been linked to the $\alpha-\mathcal{T}_3$ model exhibiting nonquantized Berry phase. Quantum interference induced localization properties of pseudospin-1 fermions om the context of the $\alpha-\mathcal{T}_3$ lattice have been recently studied by us in Ref.~\cite{singh2023quantum}, while optical conductivity in higher pseudospins has been studies in Ref.~\cite{wareham2023optical}.

%DISCUSS LOCALIZATION

%BERRY PHASE--BACKSCATTERING IN PSUEUDOSPIN

In this Letter, we solve the problem of quantum interference in \textcolor{black}{two-dimensional} fermions with arbitrary pseudospin ($s$) dispersing linearly with momentum ($\epsilon_\mathbf{k}^{ss'}\sim s' k$), where $s$ can be either a positive integer or half-integer and $-s\leq s'\leq s$, increasing in steps of unity. We derive exact analytical expressions for the fermionic wavefunctions (\textit{that we show mathematically related to Pascal's triangle}), elastic scattering time, renormalized semiclassical velocity, Cooperon, and magnetoconductivity. Evaluation of the Cooperon gaps demonstrates that weak antilocalization occurs for half-integer pseudospins, while weak localization occurs for integer pseudospins. Remarkably, we find that the gapless Cooperon mode resulting in (anti)localization behavior depends only on the pseudospin, even when multiple bands cross the Fermi energy (for $s> 1$).
Therefore, if the Fermi surface consists of multiple pockets, localization corrections from all such bands are qualitatively similar. We discover weak localization (antilocalization) behavior for integer (half-integer) pseudospin ($s$), irrespective of the band index $s'$. For flat bands, we find zero quantum correction to conductivity. A surprising facet of our analysis is that the relative localization corrections \textcolor{black}{decrease} with $s$, i.e., faster-moving relativistic electrons are \textcolor{black}{less} susceptible to disorder effects. \textcolor{black}{This result sheds insight into the physics of Anderson and many-body localization in these materials.}
Our work generalizes the past work on Weyl and Dirac fermions and provides crucial insights into the behavior of disordered electrons, paving the way for further fundamental explorations.
\begin{figure}
    \centering
    \includegraphics[width=\columnwidth]{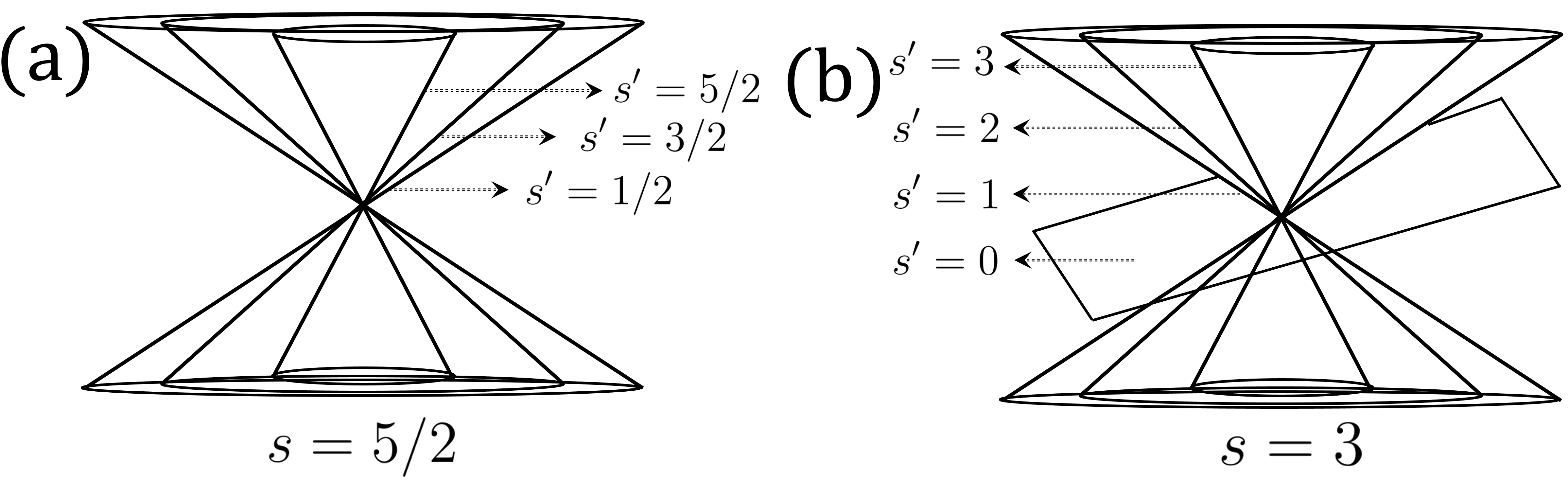}
    \caption{Energy dispersion of pseudospin-$s$ fermions plotted for two specific cases: $s=5/2$ and $s=3$.}
    \label{fig:spins}
\end{figure}

\section{Theoretical formalism}
\subsection{Hamiltonian and wavefunctions}
Pauli spin$-1/2$ matrices are generalized to the following matrices that describe fermions with pseudospin $s$:
\begin{align}
\left(S_x\right)_{\alpha\beta} & =\frac{1}{2}\left(\delta_{\alpha, \beta+1}+\delta_{\alpha+1, \beta}\right) \sqrt{(s+1)(\alpha+\beta-1)-\alpha \beta} \nonumber\\
\left(S_y\right)_{\alpha \beta} & =\frac{i }{2}\left(\delta_{\alpha, \beta+1}-\delta_{\alpha+1, \beta}\right) \sqrt{(s+1)(\alpha+\beta-1)-\alpha \beta} \nonumber\\
\left(S_z\right)_{\alpha \beta} & =(s+1-\alpha) \delta_{\alpha, \beta}=(s+1-\beta) \delta_{\alpha, \beta}
\end{align}
where 
$1 \leq \alpha \leq 2 s+1, \quad 1 \leq \beta \leq 2 s+1$, and the pseudospin $s\in \mathbb{Z}^+/2$. We consider a low-energy $k-$space Hamiltonian of the type: 
\begin{align}
H^s_\mathbf{k} = \hbar\vartheta\hspace{1mm}\mathbf{S}\cdot\mathbf{k},
\label{H_k_1}
\end{align}
where $\vartheta$ is a parameter that has dimensions of velocity, and $\mathbf{k}=(k_x,k_y)$, thus restricting ourselves to only two dimensions, although three-dimensional fermions are anticipated to exhibit qualitatively similar behavior. \textcolor{black}{It is argued that Weyl fermions in both two and three dimensions exhibit weak antilocalization, whereas Schrödinger fermions in both two and three dimensions show weak localization. However, the quantitative effects depend on the dimensionality. For instance, the velocity correction coefficient ($\eta$) is 2 for massless Weyl fermions in 2D, but 3/2 for massless Weyl fermions in 3D. Similarly, the ratio of dressed to bare Hikami boxes is $-1/4$ in for massless Weyl fermions in 2D, but $-1/6$ for massless Weyl fermions in 3D~\cite{fu2019quantum,lu2015weak}. The temperature dependence of conductivity also varies with dimensionality; however, this aspect is beyond the scope of the present study. While the chiral anomaly appears in three-dimensional Weyl fermions and is absent in two-dimensional fermions, this distinction does not impact the physics of weak localization. Strong localization (Anderson localization) differs significantly between two- and three-dimensional materials, which can be explored in future studies.
As a result, we expect certain qualitative features of weak (anti)localization to remain consistent when comparing 2D and 3D pseudospin fermions. For example, in Ref.~\cite{miao2023weak}, the authors specifically study three-dimensional spin-1 fermions and report weak localization, which agrees with our broad conclusion.
Addressing 3D massless fermions with arbitrary pseudospin is a complex problem that falls outside the scope of this manuscript, but insights from the physics of 2D fermions may provide valuable clues. Furthermore, the current work is expected to be relevant to study transport in quasi-two-dimensional thin films of proposed compounds such as  Ag$_3$Se$_2$Au, Pd$_3$Bi$_2$S$_2$,  MgPt, Li$_2$Pd$_3$B, CuBi$_2$O$_4$, and so on~\cite{bradlyn2016beyond}.  }

Eq.~\ref{H_k_1} generalizes the massless Weyl Hamiltonian and provides the low-energy theory for pseudospin-$s$ fermions with arbitrary pseudospin. Candidate materials for $s=1$ and $s=3/2$ are presented in Ref.~\cite{bradlyn2016beyond}.
The Hamiltonian above has $2s+1$ eigenvalues: $\epsilon_\mathbf{k}/(\hbar \vartheta) = \{ks, k(s-1), k(s-2),...,-ks \}$.
When $s$ is an integer, we obtain a dispersionless flat band ($\epsilon_\mathbf{k}=0$), which is absent for half-integer pseudospin (Fig.~\ref{fig:spins}). Without any loss of generality, we assume the Fermi surface to be electron-doped.  
\begin{figure}
    \centering
\includegraphics[width=\columnwidth]{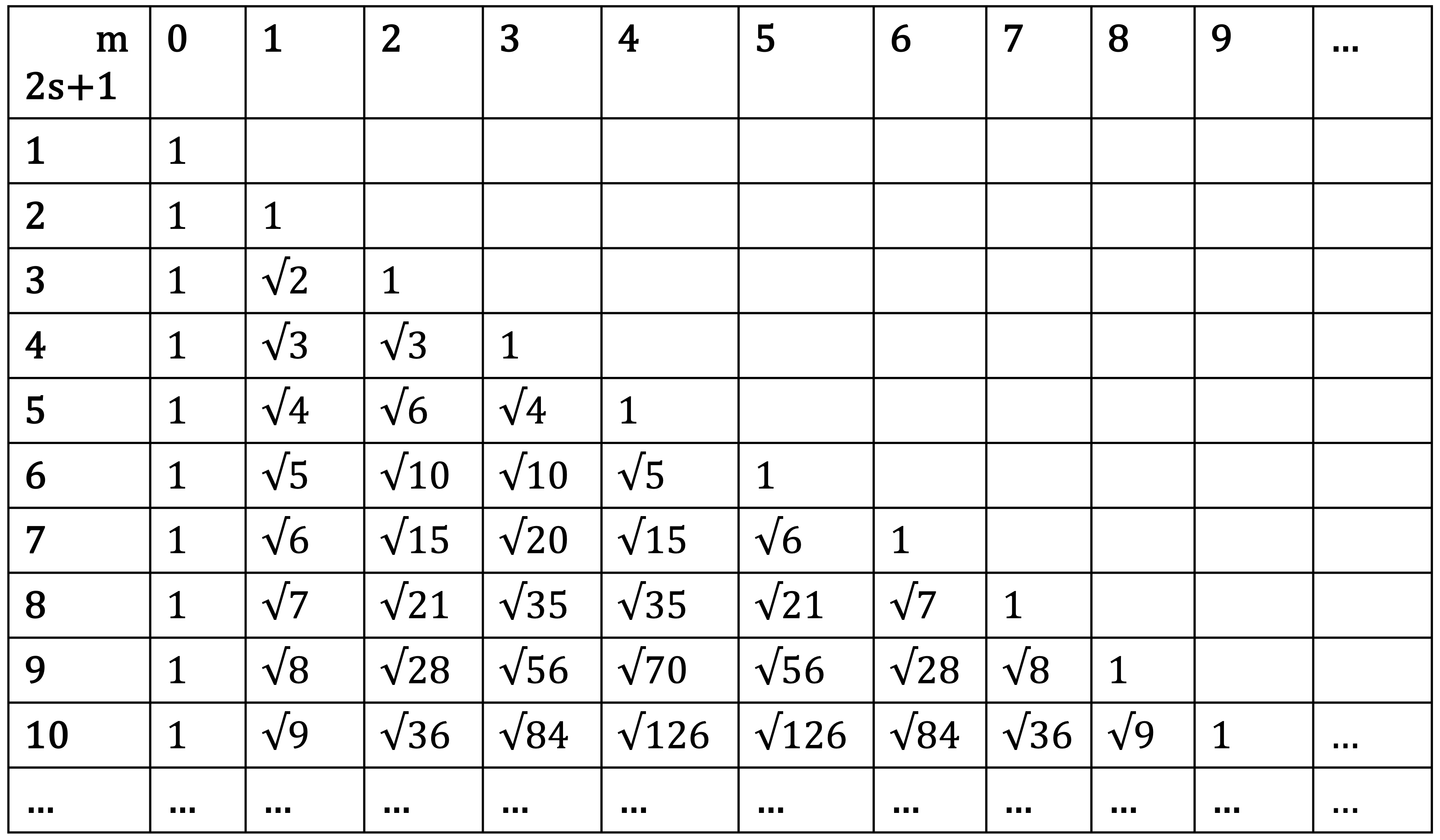}
    \caption{\textcolor{black}{Wavefunction coefficients $f_m^{ss}$ are related to the `square-root' of the entries of Pascal's triangle.}}
    \label{fig:wfcoeff}
\end{figure}
When $s\geq 3/2$, multiple bands cross the Fermi energy, and we must consider the combined effect from all such bands. Therefore, we denote the energy dispersion of any band by $\epsilon^{ss'}_\mathbf{k} = \hbar \vartheta s' k$, where the first label in the superscript $(ss')$ indicates the fermion pseudospin ($s$), and the second label indicates the particular band with dispersion $\hbar\vartheta s'k$. 
Rotational symmetry dictates the corresponding eigenfunctions to take the form
\begin{align}
    |\mathbf{k}ss'\rangle = \mathcal{N}_{ss'}\sum\limits_{m=0}^{2s} f_m^{ss'} e^{-im\phi} {|m\rangle_{s}}\color{black},
\end{align}
where $\tan\phi=k_y/k_x$, $f^{ss'}_m$ are the coefficients, $\mathcal{N}_{ss'}$ is the normalization constant, \textcolor{black}{and $|m\rangle_{s}$ is the spinor, the dimensionality of which depends on the pseudospin. For example, }
\begin{align}
&\textcolor{black}{
    |0\rangle_{1/2} = 
    \begin{pmatrix}
    1\\
    0
    \end{pmatrix};
    |1\rangle_{1/2} = 
    \begin{pmatrix}
    0\\
    1
    \end{pmatrix}}\\
    &\textcolor{black}{
    |0\rangle_{1} = 
    \begin{pmatrix}
    1\\
    0\\
    0
    \end{pmatrix};
    |1\rangle_{1} = 
    \begin{pmatrix}
    0\\
    1\\
    0
    \end{pmatrix};
    |2\rangle_{1} = 
    \begin{pmatrix}
    0\\
    0\\
    1
    \end{pmatrix}, \mathrm{etc.} }
\end{align}
\textit{Notably, we discover that the coefficients $f^{ss}_m$ have the structure of Pascal's triangle, as also shown in Fig.~\ref{fig:wfcoeff}}. \textcolor{black}{Specifically, the coefficients are evaluated to be:}
\begin{align}
&f^{ss}_m=\left(\frac{\Gamma(2s+1)}{\Gamma(m+1)\Gamma(2s+1-m)}\right)^{1/2},
\end{align}
\textcolor{black}{and the normalization constant $\mathcal{N}_{ss}= {1}/{\sqrt{2^{2s}}}$. Note that there is no assumption or restriction on the dimensionality of the pseudospin. Analytical (and even numerical) progress in systems with many degrees of freedom rapidly declines due to increasing mathematical complexity. As a result, it is often impossible to analytically evaluate the eigenfunctions and eigenenergies of large systems. This is why, despite numerous studies of weak localization in two-band systems, extensions to multiband systems have been rare.
Beyond being an intriguing aspect in its own right, the discovered connection between wavefunction coefficients and Pascal's triangle entries allows analytical calculations, which would otherwise become cumbersome or even infeasible for larger pseudospins. 
The analytical formalism now allows us to systematically evaluate all the interesting quantities like velocity renormalization, and Cooperon corrections for an arbitrary $s$, allowing us to make broad conclusions on the nature of quantum interference and localization in systems with higher $s$. It also enables us to extrapolate our results, concluding that localization corrections increase rapidly with pseudospin, as will be discussed later.
On the other hand, when $s\neq s'$, we do not find neat analytical expressions for $f^{ss'}_m$, and calculations are limited to a case-by-case basis. The discovered mathematical connection also may lead to interesting connections between Binomial coefficients and the higher irreducible representations of the SU(2) group, however, this study is beyond the scope of the current manuscript and shall be explored in future studies. }
\begin{figure}
    \centering
\includegraphics[width=\columnwidth]{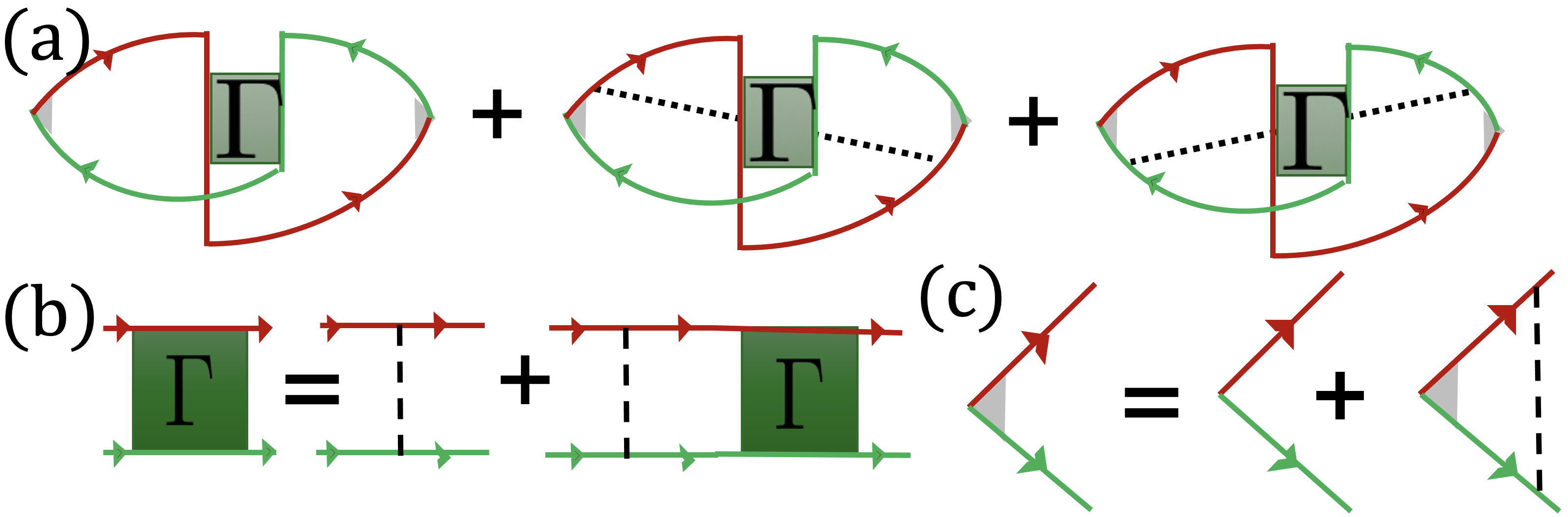}
    \caption{(a) Leading order Feynman diagrams for quantum interference correction to conductivity--bare and two dressed Hikami boxes. (b) Bethe-Salpeter equation for the Cooperon $\boldsymbol{\Gamma}$. (c) Vertex correction to the velocity.  The solid and dashed lines represent Green’s functions and impurity scattering, respectively}
    \label{fig:feyn}
\end{figure}
\begin{figure}
    \centering
    \includegraphics[width=0.9\columnwidth]{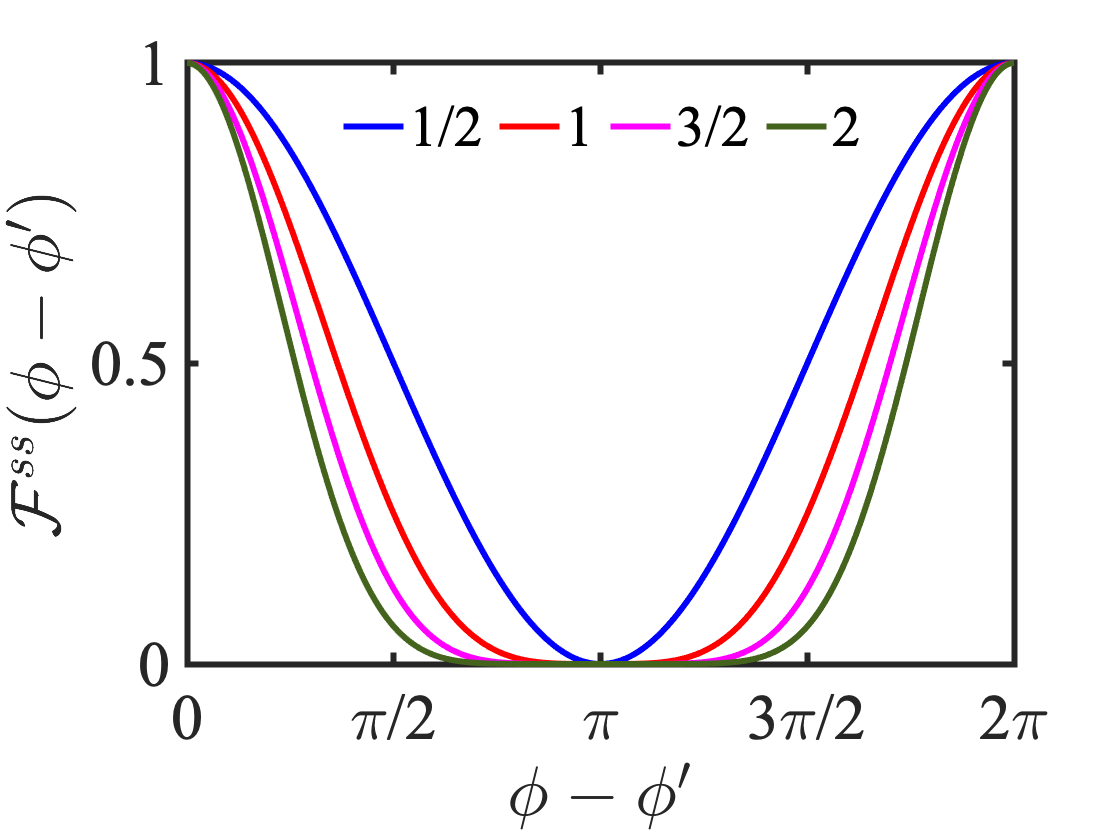}
    \caption{\textcolor{black}{The scattering probability $\mathcal{F}^{ss}(\phi-\phi')$ as a function of the incoming ($\phi$) and outgoing angle ($\phi'$). Legends indicate pseudospin $s$.}}
    \label{fig:scattprob}
\end{figure}
\subsection{Disorder and scattering time}
We consider $\delta-$correlated scalar non-magnetic impurities given by the impurity potential 
\begin{align}
U_{0}(\textbf{r})=\sum_i u_{0}\mathbb{I}_{2s+1\times 2s+1}\delta(\textbf{r}-\textbf{R}_{i}),
\label{Eq_disorder}
\end{align}
where the sum is over all impurity sites and $u_0$ is average the impurity strength. \textcolor{black}{In Eq.~\ref{Eq_disorder}, we assume that scattering by disorder does not mix the various bands. For elastic scattering, energy conservation implies that scattering occurs on the Fermi surface at sufficiently low temperatures. When $s<3/2$, only a single band crosses the Fermi level. When $s\geq 3/2$, multiple bands cross the Fermi surface, making interband elastic scattering possible. However, this involves a large momentum transfer even in the case of forward scattering, so we neglect these processes.  However, for very large pseudospins, where the bands are closely spaced, intraband scattering may play a more significant role. Since we later demonstrate that localization or antilocalization is independent of the specific band ($s'$) and depends only on the pseudospin ($s$), we do not anticipate that accounting for intraband effects will alter the fundamental physics of localization. This issue may be addressed in detail in future works. Another possibility to account for in the disorder potential is the inclusion of time-reversal-breaking (pseudo)magnetic impurities that couple to pseudospin. They may lead to spin-dependent scattering probability, or even flip the spin. On general grounds, when time-reversal symmetry is broken microscopically, the system undergoes a crossover to the unitary class, where both weak localization (WL) and weak antilocalization (WAL) are suppressed due to the cancellation paths related by time-reversal symmetry.}

The scattering (Born) amplitude is $U^{ss'}_{\mathbf{k}\mathbf{k}'}=\bra{\mathbf{k}ss'} U_{0}(\textbf{r})\ket{\mathbf{k}'ss'}$, and 
the impurity average assumes the form $\langle U^{ss'}_{\mathbf{k}\mathbf{k}'} U^{ss'}_{\mathbf{k}'\mathbf{k}} \rangle_{\mathrm{imp}} = {nu_0^2} \mathcal{F}^{ss'}(\phi-\phi')$. The scattering time calculated via the Fermi's Golden rule is 
\begin{align}
    \frac{1}{\tau_{ss'}} = \frac{2\pi}{\hbar} N^{s'}_F \mathcal{G}_{ss'} n_0u_0^2, 
    \label{Eq_tau}
\end{align}
where $N_F^{s'}={E_F}/{2\pi(s'\hbar \vartheta)^2}$ is the density of states at the Fermi energy. \textcolor{black}{The coefficients $\mathcal{G}_{ss}$ and the function $\mathcal{F}^{ss}(\phi)$for $s=s'$ take the following form:}
\begin{align}
    &\textcolor{black}{\mathcal{F}^{ss}(\phi)= \cos^{4s}(\phi/2)}\nonumber\\
    &\textcolor{black}{\mathcal{G}_{ss}= \frac{\Gamma(2s+1/2)}{\sqrt{\pi}\Gamma(2s+1)}},
\end{align}
\textcolor{black}{while coefficients $\mathcal{G}_{ss'}$  for $s\neq s'$ is specifically evaluated in the Appendix E.} 
\textcolor{black}{In Fig.~\ref{fig:scattprob}, we plot the scattering probability $\mathcal{F}^{ss}(\phi-\phi')$ as a function of incoming and outgoing angle. When $s=1/2$, perfect backscattering is forbidden, i.e., $\mathcal{F}^{ss}(\phi-\phi'=\pi)=0$ as expected. This increases transport time compared to the scattering time $\tau_{ss}$ as reported in graphene~\cite{mccann2006weak}. With increasing $s$, a substantial region around $\phi-\phi'=\pi$ shows suppressed scattering probability. Consequently, the transport time compared to the scattering time is much more enhanced with increasing $s$. This renormalization is also reflected in the velocity vertex correction as discussed next. }

\subsection{Velocity correction and the Bethe-Salpeter equation}
\textcolor{black}{Unlike Schr\"odinger fermions, fermions described in our model have a momentum-independent current vertex. The velocity vertex that enters the conductivity therefore is renormalized. }
We therefore need to evaluate the ladder diagram correction to the quasiclassical velocity. The corresponding equation is given by (Fig.~\ref{fig:feyn} (c))
\begin{equation}
\tilde{\mathbf{v}}_{\mathbf{k}}^{ss'}=\mathbf{v}_{\mathbf{k}}^{ss'}+\sum_{\mathbf{k}^{\prime}} G^{ss'R}_{\mathbf{k}^{\prime}} G^{ss'A}_{\mathbf{k}^{\prime}}\left\langle U^{ss'}_{\mathbf{k}\mathbf{k}^{\prime}} U^{ss'}_{\mathbf{k}^{\prime}\mathbf{k}}\right\rangle_\mathrm{{imp}}\tilde{\mathbf{v}}_{\mathbf{k}^{\prime}}^{ss'},
\label{Eq_vel_1}
\end{equation}
where $\tilde{\mathbf{v}}_{\mathbf{k}}^{ss'}$ and ${\mathbf{v}}_{\mathbf{k}}^{ss'}$ denote the impurity-dressed and bare velocity, respectively. $G^{ss'R}_{\mathbf{k}^{\prime}}$ and $G^{ss'A}_{\mathbf{k}^{\prime}}$ are retarded and advanced Green's functions, respectively, and are given by 
\begin{align}
    G^{ss'R/A}_{\mathbf{k}} (\omega)= \frac{1}{\omega-\epsilon^{ss'}_\mathbf{k} \pm \frac{i\hbar}{2\tau_{ss'}}}
\end{align}
The ansatz $\tilde{\mathbf{v}}_{\mathbf{k}}^{ss'}= \eta^{ss'}\mathbf{v}_{\mathbf{k}}^{ss'}$ exactly solves Eq.~\ref{Eq_vel_1}, and $\eta^{ss'}$ is evaluated \textcolor{black}{in Appendix E}. Notably, we find 
\begin{align}
\eta^{ss}=2s+1. 
\end{align}
\textcolor{black}{The renormalization correction of the vertex $\mathbf{v}_\mathbf{k}^{ss}$ increases with pseudospin $s$. On the other hand, the correction of the vertex $\mathbf{v}_\mathbf{k}^{ss'}$ for a fixed $s'$ decreases with $s$ (see Table~\ref{tab:etasspO} in Appendix E ). }
The quantum interference correction to conductivity, obtained by summing the contribution of a bare Hiakmi box ($\sigma_0^F$) and two dressed Hikami boxes ($\sigma_A^F$ and $\sigma_A^R$, (Fig.~\ref{fig:feyn} (a))). 
\textcolor{black}{The bare Hikami box at zero temperature is calculated as}
\begin{equation}
\textcolor{black}{
     \sigma_{0}^{F}=\frac{e^{2}\hbar}{2\pi}\sum_{\textbf{q}}\boldsymbol{\Gamma}(\textbf{q})\sum_{\textbf{k}}\Tilde v_{\textbf{k}}^{ss'x}\Tilde v^{ss'x}_{\textbf{q-k}}G_{\mathbf{k}}^{ss'\mathrm{R}} G_{\mathbf{k}}^{ss'\mathrm{A}}G_{\mathbf{q-k}}^{ss'\mathrm{R}} G_{\mathbf{q-k}}^{ss'\mathrm{A}},}
\end{equation}
\textcolor{black}{The two dressed Hikami boxes are calculated as} 
\begin{align}
\textcolor{black}{
    \sigma_{R}^{F}= \frac{e^{2}\hbar}{2\pi}\sum_{\textbf{q}}}&\textcolor{black}{\Gamma(\textbf{q})\sum_{\textbf{k}}\sum_{\textbf{k}_{1}}\Tilde v_{\textbf{k}}^{ss'x}\Tilde v^{ss'x}_{\textbf{q}-\textbf{k}_{1}}G_{\mathbf{k}}^{ss'\mathrm{R}}G_{\mathbf{k_1}}^{ss'\mathrm{R}}G_{\mathbf{q-k}}^{ss'\mathrm{R}}}\nonumber\\
    &\textcolor{black}{G_{\mathbf{q-k_1}}^{ss'\mathrm{R}}G_{\mathbf{k}}^{ss'\mathrm{A}} G_{\mathbf{q-k_1}}^{ss'\mathrm{A}}\langle U^{ss'}_{\mathbf{k}_{1},\textbf{k}}U^{ss'}_{\mathbf{q-k_{1}},\mathbf{q-k}}\bigl \rangle_{\mathrm{imp}}},\nonumber\\
         \textcolor{black}{\sigma_{A}^{F}= \frac{e^{2}\hbar}{2\pi}\sum_{\textbf{q}}}&\textcolor{black}{\Gamma(\textbf{q})\sum_{\textbf{k}}\sum_{\textbf{k}_{1}}\Tilde v_{\textbf{k}}^{ssx}\Tilde v^{ssx}_{\textbf{q}-\textbf{k}_{1}}G_{\mathbf{k}}^{ss\mathrm{R}}G_{\mathbf{q-k_1}}^{ss\mathrm{R}}G_{\mathbf{k}}^{ss\mathrm{A}}}\nonumber\\
     &\textcolor{black}{G_{\mathbf{k_1}}^{ss\mathrm{A}}G_{\mathbf{q-k}}^{ss\mathrm{A}} G_{\mathbf{q-k_1}}^{ss\mathrm{A}}\langle U^{ss}_{\textbf{k},\mathbf{k}_{1}}U^{ss}_{\mathbf{q-k},\mathbf{q-k_1}}\bigl \rangle_{\mathrm{imp}},}
\end{align}
\textcolor{black}{In the small-$q$ limit, we find:}
\begin{align}
    \sigma_0^F &= -\frac{e^2 {s'}^2 \vartheta^2 N^{s'}_F \eta_{ss'}^2\tau_{ss'}^3}{\hbar^2} \sum_{\textbf{q}}\boldsymbol{\Gamma}(\textbf{q});\quad \sigma_A^R=\sigma_A^F;\nonumber\\
    \sigma_{F}^{A} &=\frac{e^2 N^{s'}_F \tau_{ss'}^3\eta_{ss'}^2\vartheta^2{s'}^2 }{4\hbar^2 \mathcal{G}_{ss'}} \mathcal{A}^{ss'}_1  \sum_{\textbf{q}}\boldsymbol{\Gamma}(\textbf{q}),
    \label{Eq:Cond}
\end{align}
where $\boldsymbol{\Gamma} (\mathbf{q})$ is the vertex ((Fig.~\ref{fig:feyn} (b))), and $\mathcal{A}^{ss'}_m$ are the coefficients of the bare vertex, defined in Eq.~\ref{Eq_Gamma0}. As a sanity check, we recover the results for graphene~\cite{suzuura2002crossover,khveshchenko2006electron,mccann2006weak}: 
\begin{align}
\eta^{\frac{1}{2}\frac{1}{2}}=2; \mathcal{F}^{\frac{1}{2}\frac{1}{2}}(\phi) = \cos^2(\phi/2); {\sigma_{A}^{F}}/{\sigma_{0}^{F}} = -{1}/{4}
\end{align}
The Bethe-Salpeter equation for the vertex is given by
\begin{align}
    \boldsymbol{\Gamma}^{ss'}_{\mathbf{k}_{1}, \mathbf{k}_{2}}= \boldsymbol{\Gamma}_{\mathbf{k}_{1}, \mathbf{k}_{2}}^{ss'0}+\sum_{\mathbf{k}}\boldsymbol{\Gamma}_{\mathbf{k}_{1}, \mathbf{k}}^{ss'0} G_{\mathbf{k}}^{ss'R}G_{\mathbf{q}-\mathbf{k}}^{ss'A} \boldsymbol{\Gamma}^{ss'}_{\mathbf{k}, \mathbf{k}_{2}},
    \label{Eq_bethe}
\end{align}
where the bare vertex $
\boldsymbol{\Gamma}^{ss'0}_{\mathbf{k_{1},k_{2}}}= \langle U^{ss'}_{\mathbf{k_1}\mathbf{k}_{2}}U^{ss'}_{\mathbf{-k_1}\mathbf{k_2}}\bigl\rangle_{\mathrm{imp}}$ is evaluated to
\begin{align}
\boldsymbol{\Gamma}^{ss'0}_{\mathbf{k_{1},k_{2}}} = \left(\frac{\hbar}{2\pi N^{s'}_F \mathcal{G}_{ss'} \tau_{ss'}}\right) \sum\limits_{m=0}^{4s} {\mathcal{A}^{ss'}_m} e^{im(\phi_1-\phi_2)}.
\label{Eq_Gamma0}
\end{align}
\textcolor{black}{For $s=s'$, we analytically evaluate the coefficients for arbitrary pseudospin:}
\begin{align}
    &\textcolor{black}{
    \mathcal{A}^{ss}_{0\leq m\leq 2s}= \frac{\Gamma(2s+1/2)}{\sqrt{\pi} \left(\prod\limits_{k=0}^{k=2s-m-1}{4s-m-k}\right) m!}}\nonumber\\
    &\textcolor{black}{\mathcal{A}^{ss}_{2s\leq m\leq 4s} =\mathcal{A}^{ss}_{4s-m}}.
\end{align}
\textcolor{black}{The coefficients $\mathcal{A}^{ss'}_m$ for $s\neq s'$ are specified explicitly in the Appendix E}. 

\begin{figure*}
    \centering
\includegraphics[width=1.8\columnwidth]{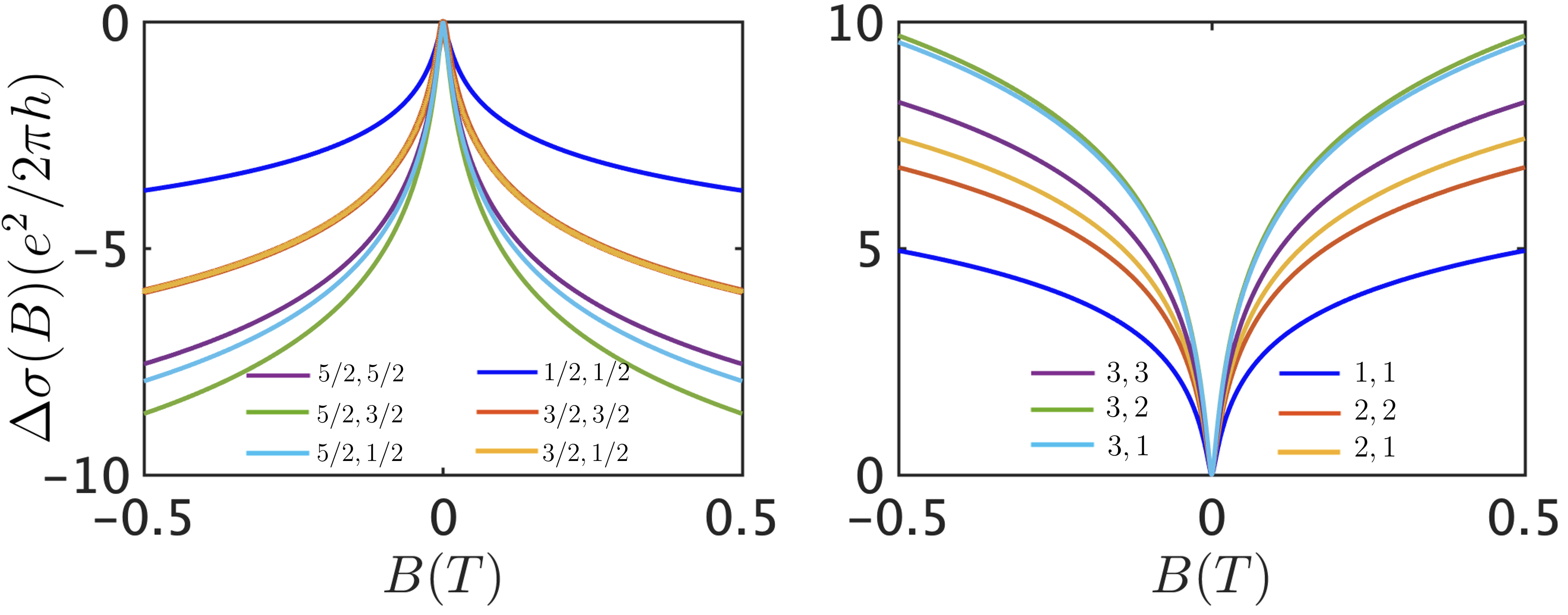}
    \caption{Mangetoconductivity of pseudospin-$s$ fermions. (a) WAL behavior for half-integer $s$. (b) WL behavior for integer $s$.  We choose $l_\phi$=300 nm. Legends indicate $\{s,s'\}$.}
    \label{fig:delsigma}
\end{figure*}
\begin{figure}
    \centering
    \includegraphics[width=0.8\columnwidth]{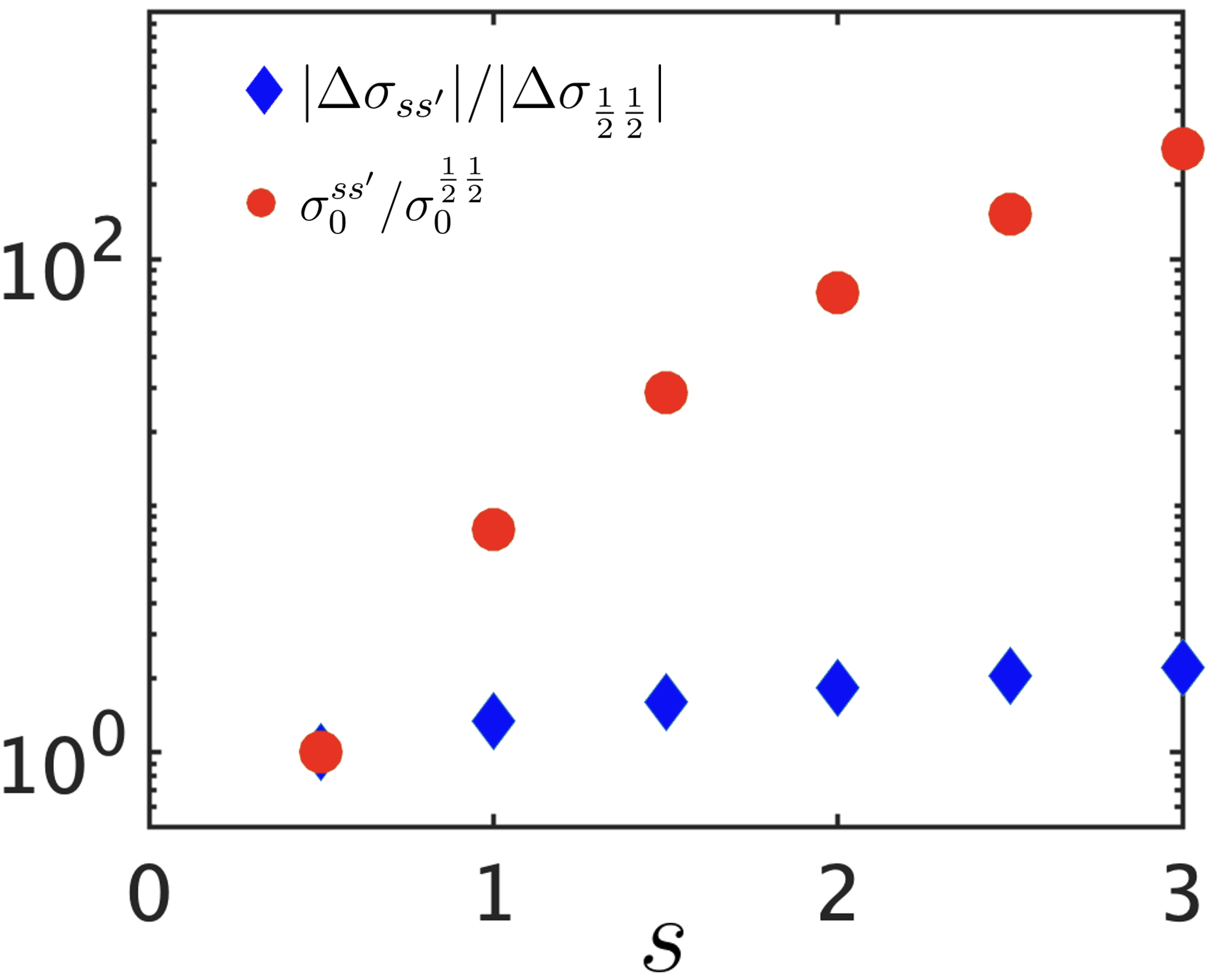}
    \caption{The relative increase of localization induced magnetoconductivity (blue) and the Drude conductivity (red) for pseudospin-$s$ fermions. We choose $s'=s$. }
    \label{fig:reldelsigma}
\end{figure}
\begin{figure}
    \centering
    \includegraphics[width=0.8\linewidth]{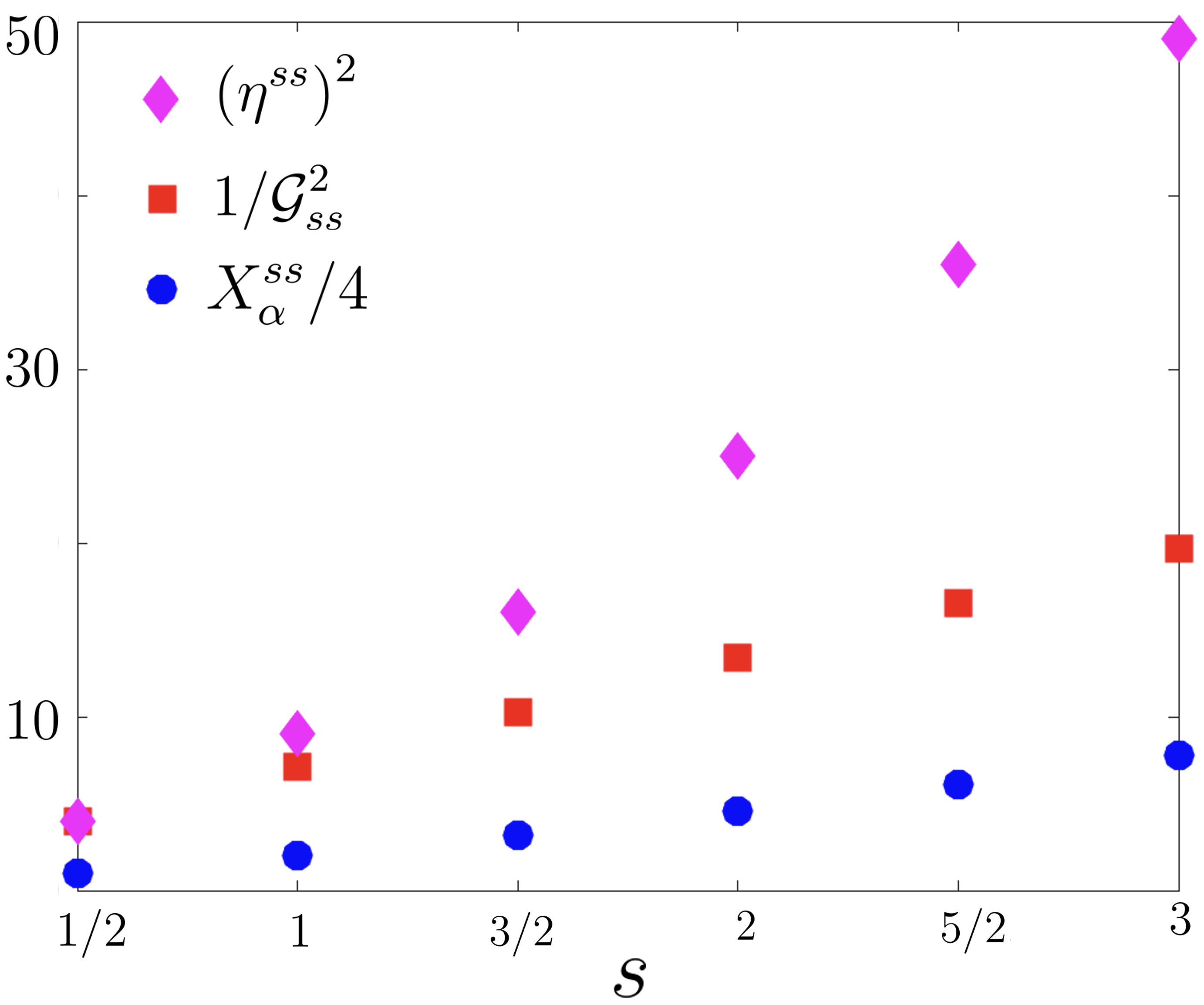}
    \caption{\textcolor{black}{The coefficients $(\eta^{ss})^2$, $1/\mathcal{G}_{ss}^2$, $X_\alpha^{ss}/4$ as a function of pseudospin-$s$.}}
    \label{fig:eta_X_G}
\end{figure}

To solve for the dressed vertex, we choose the following ansatz:
\begin{align}
\boldsymbol{\Gamma}^{ss'}_{\mathbf{k_{1},k_{2}}} = \left(\frac{\hbar}{2\pi N^{s'}_F \mathcal{G}_{ss'} \tau_{ss'}}\right) \sum\limits_{m=0}^{4s}\sum\limits_{n=0}^{4s} {\mathcal{V}^{ss'}_{mn}} e^{i(m\phi_1-n\phi_2)}.
\label{Eq_V_ansatz}
\end{align}
This solves the Bethe-Salpeter equation Eq.~\ref{Eq_bethe}. The coefficients of the matrix $\mathcal{V}^{ss'}$ are given by the solution of the following equation: 
\begin{align}{\mathcal{V}}^{ss'} = (1-\mathcal{A}^{ss'} \Phi^{ss'}\mathcal{G}_{ss'}^{-1})^{-1} \mathcal{A}^{ss'},
\label{Eq_V_matrix}
\end{align}
where 
\begin{align}
    \Phi^{ss'}_{mn} &= \int{\frac{d\phi}{2\pi} \frac{e^{i(n-m)\phi}}{1+i\tau_{ss'} \vartheta s' q \cos\phi}}\nonumber\\
    &= \left(1-\frac{Q^2}{2}\right) \delta_{mn} -\frac{iQ}{2}\left(\delta_{m,n+1}+\delta_{m,n-1}\right) \nonumber\\
    &-\frac{Q^2}{4}\left(\delta_{m,n+2}+\delta_{m,n-2}\right),
\end{align}
and $Q=\vartheta\tau_{ss'} s'q$.
The diverging elements of $\mathcal{V}^{ss'}$ give us information about the vanishing Cooperon gaps that are significant to understanding the localization behavior. 
\textcolor{black}{It is possible to express the diagonal elements of the matrix $\mathcal{V}^{ss'}$ as (see Appendix C)}
\begin{align}
\textcolor{black}{
\mathcal{V}^{ss'}_{ii} = \frac{\mathcal{G}_{ss'}}{\left(-1+\frac{\mathcal{G}_{ss'}}{\mathcal{A}^{ss'}_{i}}\right)+\mathcal{S}^{ss'}_{ii} \frac{Q^2}{\prod_i\left(-1+\frac{\mathcal{G}_{ss'}}{\mathcal{A}^{ss'}_i}\right)}},}
\end{align}
\textcolor{black}{where $\mathcal{S}$ is defined in the Appendix C. When $(-1+\mathcal{G}_{ss'}/\mathcal{A}^{ss'}_\alpha)\rightarrow 0$, the elements $\mathcal{V}^{ss'}_{\alpha\alpha}$ diverge in the limit $q\rightarrow 0$. We therefore term $(g^{ss'}_\alpha\equiv 2(-1+\mathcal{G}_{ss'}/\mathcal{A}^{ss'}_\alpha))$ as the Cooperon gaps. Therefore in the limit $q\rightarrow 0$, the vertex correction is dominated by the following term:}
\begin{align}
    \textcolor{black}{\boldsymbol{\Gamma}^{ss'}_{\mathbf{k}_1\mathbf{k}_2} \sim  \frac{1}{q^2}e^{i\alpha(\phi_1-\phi_2)}.}
\end{align}
\textcolor{black}{In the vertex, $\mathbf{k}_2 = \mathbf{q}-\mathbf{k}_1\approx \mathbf{k}_1$, then $\phi_2=\pi+\phi_1$.}

\section{Conductivity}
The zero-field quantum interference correction to the conductivity from the gapless Cooperon mode $\alpha$ (for the band $|\mathbf{k}ss'\rangle$) is evaluated by summing the Hikami boxes:
\begin{align}
    \sigma_{ss'} = -\frac{e^2}{2\pi h}  {Y^{ss'}_\alpha}\ln(l_\phi/l_{ss'})e^{i\alpha\pi},
    \label{Eq:sigma_1}
\end{align}
where $l_\phi$ is the coherence length, and
\begin{align}
    & Y^{ss'}_\alpha = \frac{{\eta^{ss'}}^{2} }{2 X^{ss'}_\alpha {\mathcal{G}^{ss'}}^2} \left(1-\frac{\textcolor{black}{\mathcal{A}^{ss'}_{\alpha-1}+\mathcal{A}^{ss'}_{\alpha+1}}}{2\mathcal{G}^{ss'}}\right), 
   \nonumber\\
    & {l_{ss'}}^{-2} = \frac{2}{\vartheta^2 \tau_{ss'}^2},\quad  {X}_\alpha^{ss'}= \frac{2}{\mathcal{V}_{\alpha\alpha}^{ss'} }
    \label{Eq:coeffs2}
\end{align}
Remarkably, we discover that the gapless Cooperon mode $\alpha$ is independent of the band index $s'$ and only depends on the pseudospin $s$~\cite{SI}. Specifically, we find 
\begin{align}
\alpha = 2s.
\label{Eq_alpha}
\end{align}
Therefore, even if multiple bands ($|\mathbf{k}ss'\rangle$ and $|\mathbf{k}ss''\rangle$) intersect the Fermi energy, localization corrections from all of them are qualitatively similar. We find that for half-integer (integer) pseudospin, $e^{i\alpha\pi}=-1$ ($e^{i\alpha\pi}=+1$), resulting in weak antilocalization (localization) behavior. \textcolor{black}{Recall that the Berry phase for $|\mathbf{k}ss'\rangle$ is $e^{2\pi is'}$.
Therefore, from Eq.~\ref{Eq_alpha}, the exponential factor $e^{i\alpha\pi}$ can also be identified with the Berry phase of the pseudospin. Interestingly, it is the Berry phase of the pseudospin that enters the equation (Eq.~\ref{Eq:sigma_1}) and not the Berry phase of the particular band, but since they are identical ($e^{2\pi is}=e^{2\pi is'}$ for a given pseudospin $s$, if $s'\neq 0$,} \textcolor{black}{and the fact that Berry phase is defined modulo $2\pi$) in this model, it does not lead to any measurable difference.} \textcolor{black}{Physically, this is understood as follows: during backscattering, the pseudospin rotates by $\pi \alpha$ resulting in a phase difference between the two interfering paths. Whether the spin rotates by $\pi$($2\pi$) or by $3\pi$($4\pi$), the phase factor (-1 or +1) remains the same in both cases. Interband scattering effects are also not likely to change this qualitative behavior as discussed in Sec. IIA.
Note that even though the Berry phase contribution is independent of $s'$, the numerical factor $Y_\alpha^{ss'}$ depends on $s'$, and thus the conductivity corrections for $|\mathbf{k}ss'\rangle$ and $|\mathbf{k}ss''\rangle$ are quantitatively different even though they are qualitatively similar.}
We also predict that for flat bands ($s'=0$), quantum corrections vanish. 

With the application of a magnetic field, the phase coherence is lost and the quantum correction is suppressed, enabling experimental observation of weak localization and weak antilocalization corrections through magnetoconductivity measurements~\cite{bergmann1984weak}. This is derived by quantizing the wavevector $q^2\rightarrow (n+1/2)(4eB/\hbar^2)$. In the weak-field limit, the magnetoconductivity ($\Delta \sigma (B)_{ss'} = \sigma(B)_{ss'} - \sigma_{ss'}$) is given by 
\begin{align}
    \Delta \sigma(B)_{ss'} = \frac{e^2}{\pi h}  Y^{ss'}_\alpha \bigg[ \Psi\left(\frac{l_B^2}{l_\phi^2} +\frac{1}{2} \right)- \ln\left(\frac{l_B^2}{l_\phi^2} \right)\bigg]e^{i\alpha\pi},
    \label{Eq:delsigmaB}
\end{align}
where $\Psi(x)$ is the digamma function. Notably, the zero-field conductivity correction (Eq.~\ref{Eq:sigma_1}) and the magnetoconductivity crucially depend on the same prefactor $Y_\alpha^{ss'}$ that governs the magnitude of the correction. 
Eq.~\ref{Eq:sigma_1}-\ref{Eq:delsigmaB} are the main results of this paper that generalize the existing results for the Dirac/Weyl fermion to arbitrary pseudospin-$s$. 

In Fig.~\ref{fig:delsigma} we plot the magnetoconductivity for both half-integer and integer pseudospin-$s$ fermions limiting ourselves to $s\leq 3$, including all $0<s'\leq s$. Remarkably, both the WAL correction (for half-integer $s$) and WL correction (for integer $s$) increase with increasing $s$. 
%On the other hand, magnetoconductivity for the same $s'$ but different $s$ have comparable orders of magnitude. For example $\{s,s'\} =\{1/2,1/2\}$, $\{3/2,1/2\}$ and $\{5/2,1/2\}$ have a similar order of magnitude, but $\{1/2,1/2\}$, $\{3/2,3/2\}$ and $\{5/2,5/2\}$ are strikingly different. Therefore the magnitude of the localization correction is strongly dependent on $s'$ and not $s$, but since larger values of $s'$ are only possible for larger values of $s$, we state that higher pseudospins lead to stronger localization correction. It can be argued that for large $s$, perturbation theory may break down at comparatively lesser magnetic fields. Nevertheless, quantum effects will still lead to strong (anti)localization that may be probed nonperturbatively in future studies. 
\textcolor{black}{The increase in the conductivity correction can be ascribed to the following factors that increase with $s$: (i) increase in the velocity vertex renormalization $\eta^{ss}$, (ii) increase in the Cooperon vertex factor $X_\alpha^{ss}$, (iii) increase in $1/\mathcal{G}_{ss}$, which is related to the inverse of the scattering matrix element (see Eq.~\ref{Eq_SI_onebytau} in Appendix A). Fig.~\ref{fig:eta_X_G} shows how these factors scale with the pseudospin $s$. These factors deviate from unity only in the case of relativistic fermions, and are absent in ordinary Schro\"{o}dinger fermions~\cite{mccann2006weak}. Furthermore, here we show that these renormalization corrections increase with the pseudospin, and collectively result in enhanced conductivity corrections. }

The  Drude conductivity for $|\mathbf{k}ss'\rangle$ can be evaluated as 
\begin{align}
    \textcolor{black}{\sigma_0^{ss'} = {e^2}\int{\frac{d^2k}{(2\pi)^2} \tau_{ss'}(v^{ss'}_x)(\tilde{v}^{ss'}_x) \left(-\frac{\partial f_0}{\partial \epsilon}\right)},} 
\end{align}
\textcolor{black}{where $f_0$ is the Fermi-Dirac distribution function. }This yields a rather simple expression at $T\rightarrow 0$:
\begin{align}
    \sigma_0^{ss'} = \textcolor{black}{\frac{e^2h}{4\pi^2}}\bigg(\frac{\vartheta^2 {s'}^2\textcolor{black}{\eta^{ss'}}}{\mathcal{G}_{ss'} n_0u_0^2}\bigg),
\end{align}
\textcolor{black}{where we have used Eq.~\ref{Eq_tau}.}
The conductivity scales approximately with the second power of $s'$. \textcolor{black}{This can be contrasted with the magnetoconductivity that doesn't explicitly scale with $s'$.}
We further compare the relative increase of the Drude conductivity and the quantum interference correction. In Fig.~\ref{fig:reldelsigma} we plot the relative increase in magnetoconductivity $|\Delta\sigma_{ss}|/|\Delta\sigma_{\frac{1}{2}\frac{1}{2}}|$ and the relative increase in the Drude conductivity $\sigma_0^{ss}/\sigma_0^{\frac{1}{2}\frac{1}{2}}$. While $\sigma_0\sim s^2$, $\Delta\sigma_{ss}$ scales \textcolor{black}{less} drastically. \textcolor{black}{The quadratic rise in Drude conductivity as a function of pseudospin compared to the logarithmic rise in the magnetoconductivity corrections imply that the quantum corrections are less dominant for higher pseudospin fermions. }

A brief note about electron-electron interactions is in order. The interaction parameter $r_s$ represents the ratio of the average inter-electron Coulomb interaction energy to the Fermi energy. The average Coulomb energy is $\langle V \rangle \sim {e^2}/{\langle r \rangle}$, where $\langle r \rangle = n^{-1/2} \sim s'/k_F$ is the average inter-particle separation. Therefore, $r_s\sim 1/s'$, which indicates that electron-electron interactions are less dominant for higher pseudospins. \textcolor{black}{As we discussed before, higher pseudospin fermions are also less susceptible to disorder effects. These two findings may have important implications in the Anderson-Mott transition~\cite{ma2018localization}, many-body localization that may be explored in upcoming studies.}

\section{Outlook}
Advances in material science have enabled the realization of a manifold of emergent electronic excitations, from massless Dirac and Weyl excitations to flat bands in moir\'{e} materials. Combined with theoretical predictions of realizing materials that host higher pseudospin fermions in solids (at least up to $s=2$~\cite{bradlyn2016beyond}), these developments open up exciting possibilities for studying quantum transport to its fullest essence. Here, we solve the fundamental problem of disorder-induced quantum interference corrections leading to electron (anti)localization in fermionic excitations that carry an arbitrary pseudospin $s$. First, we establish a mathematical connection of the wavefunctions to Pascal's triangle. Second, deriving exact analytical expressions for the relevant transport quantities reveals that the gapless Cooperon modes depend exclusively on the pseudospin, resulting in weak localization (antilocalization) behavior for integer (half-integer) $s$, irrespective of the band index. An important finding of our work is that the localization correction \textcolor{black}{increases} with increasing $s$. \textcolor{black}{Enhancement of several renormalization factors, such as the current vertex, Cooperon contributions, and transport time, collectively result in the increase of conductivity corrections with $s$. On the other hand the relative increase of localization corrections compared to the increase in Drude conductivity is found to be rather small, which implies that higher pseudospin fermions are less susceptible to disorder effects. Our elementary analysis also suggest that electron-electron interactions are also less dominant in higherpseudospin fermions. }
We generalize existing works on localization effects in Weyl and Dirac fermions and provide insights, pushing our fundamental understanding of how disorder acts in these materials. Our study will likely spur further fundamental studies on pseudospin-$s$: (i) how does conductivity scale with the system size? (ii) how do electron-electron interactions and disorder interplay?

\textit{Acknowledgement:} This work was supported by ANRF-SERB Core Research Grant CRG/2023/005628. We thank Arpan Gupta for checking the calculations. 
\begin{widetext}
\appendix

\section{Model and formalism}
\subsection{Pseudospin-$s$ fermions}
Pauli spin$-1/2$ matrices are generalized to the following matrices that describe fermions with pseudospin $s$:
\begin{align}
\left(S_x\right)_{\alpha\beta} & =\frac{1}{2}\left(\delta_{\alpha, \beta+1}+\delta_{\alpha+1, \beta}\right) \sqrt{(s+1)(\alpha+\beta-1)-\alpha \beta} \nonumber\\
\left(S_y\right)_{\alpha \beta} & =\frac{i }{2}\left(\delta_{\alpha, \beta+1}-\delta_{\alpha+1, \beta}\right) \sqrt{(s+1)(\alpha+\beta-1)-\alpha \beta} \nonumber\\
\left(S_z\right)_{\alpha \beta} & =(s+1-\alpha) \delta_{\alpha, \beta}=(s+1-\beta) \delta_{\alpha, \beta}
\end{align}
where 
$1 \leq \alpha \leq 2 s+1, \quad 1 \leq \beta \leq 2 s+1$, and the pseudospin $s\in \mathbb{Z}^+/2$. We consider a low-energy $k-$space Hamiltonian of the type: 
\begin{align}
H^s_\mathbf{k} = \hbar\vartheta\hspace{1mm}\mathbf{S}\cdot\mathbf{k},
\end{align}
where $\vartheta$ is a parameter that has dimensions of velocity. 
The Hamiltonian has $d\equiv 2s+1$ eigenvalues: $\epsilon_\mathbf{k}/(\hbar \vartheta) = \{ks, k(s-1), k(s-2),...,-ks \}$.
When $s$ is an integer, we obtain a dispersionless flat band ($\epsilon_\mathbf{k}=0$), which is absent for half-integer pseudospin. Without any loss of generality, we assume the Fermi energy to have a finite positive value (electron doping).   

When $s\geq 3/2$, multiple bands cross the Fermi energy, and we need to consider the combined effect from all those bands. 
We denote the energy dispersion of the bands by $\epsilon^{(ss')}_\mathbf{k} = +\hbar \vartheta s' k$, where the first label in the superscript $(ss')$ indicates the fermion pseudospin $s$ and the second label indicates the band with dispersion $\hbar\vartheta s'k$. 
\subsection{Generalized eigenfunctions}
The eigenfunctions corresponding to $\epsilon^{ss'}_\mathbf{k}$ take the following form
\begin{align}
    |\mathbf{k}ss'\rangle = \mathcal{N}_{ss'}\sum\limits_{m=0}^{2s} f_m^{ss'} e^{-im\phi},
    \label{Eq_wf1}
\end{align}
where $\tan\phi=k_y/k_x$, $f^{ss'}_m$ are the coefficients, and $\mathcal{N}_{ss'}$ is the normalization constant. In later sections, we provide the analytical form of $f_{m}^{ss'}$ for a few cases and the exact analytical form for $f_{m}^{ss}$. 
\subsection{Impurity potential}
We consider $\delta-$correlated scalar non-magnetic impurities given by the impurity potential 
\begin{align}U_{0}(\textbf{r})=\sum_i u_{0}\mathbb{I}_{2s+1\times 2s+1}\delta(\textbf{r}-\textbf{R}_{i}),\end{align}
where the sum is over all impurity sites and $u_0$ is the impurity strength, assumed to be the same at each site. The scattering (Born) amplitude is \begin{align}U^{ss'}_{\mathbf{k}\mathbf{k}'}=\bra{\mathbf{k}ss'} U_{0}(\textbf{r})\ket{\mathbf{k}'ss'},\end{align} and 
the impurity assumes the form 
\begin{align}\langle U^{ss'}_{\mathbf{k}\mathbf{k}'} U^{ss'}_{\mathbf{k}'\mathbf{k}} \rangle_{\mathrm{imp}} = {nu_0^2} \mathcal{F}^{ss'}(\phi-\phi'),\end{align} 
where the expression for $\mathcal{F}^{ss'}(\phi)$ will be provided later. Since the energy dispersion $\epsilon_\mathbf{k}^{ss'}$ depends only on $s'$, the density of states also depends only on $s'$ and is independent of $s$: 
\begin{align}
    N^{ss'}(E) &= \frac{1}{4\pi^2} \int\limits_0^\infty {k dk} \int\limits_0^{2\pi} d\phi {\delta(\epsilon^{ss'}_\mathbf{k}-E)}\nonumber\\
    &=\frac{1}{2\pi} \int\limits_0^\infty {dk k \delta(\epsilon^{ss'}_\mathbf{k}-E)}\nonumber\\
    &=\frac{E}{2\pi(s'\hbar \vartheta)^2}\equiv N^{s'}(E).
\end{align}
The scattering time calculated via the Fermi's Golden rule is 
\begin{align}
    \frac{1}{\tau_{ss'}}&=\frac{2\pi}{\hbar}\sum_{\textbf{k}'}\bigr \langle U^{ss'}_{\textbf{k},\textbf{k}'}U^{ss'}_{\textbf{k}',\textbf{k}}\bigl \rangle_{imp}\delta(E_{F}-\epsilon_{\textbf{k}'})\nonumber\\
        &=\frac{2\pi}{\hbar}N^{s'}_F \int\limits_0^{2\pi} \frac{d\phi'}{2\pi}\langle U^{ss'}_{\mathbf{k}\mathbf{k}'} U^{ss'}_{\mathbf{k}'\mathbf{k}} \rangle_{\mathrm{imp}} \nonumber\\
    &=\frac{2\pi}{\hbar} N^{s'}_F \mathcal{G}_{ss'} n_0u_0^2,
    \label{Eq_SI_onebytau}
\end{align} 
where $N_F^{s'}={E_F}/{2\pi(s'\hbar \vartheta)^2}$ is the density of states at the Fermi energy. The coefficient $\mathcal{G}_{ss'}$, which is obtained by the angular integration of $\langle U^{ss'}_{\mathbf{k}\mathbf{k}'} U^{ss'}_{\mathbf{k}'\mathbf{k}} \rangle_{\mathrm{imp}}$ will be specified later. 
\subsection{Velocity correction}
Next, we evaluate the ladder diagram correction to the velocity. The corresponding equation is given by 
\begin{equation}
\tilde{\mathbf{v}}_{\mathbf{k}}^{ss'}=\mathbf{v}_{\mathbf{k}}^{ss'}+\sum_{\mathbf{k}^{\prime}} G^{ss'R}_{\mathbf{k}^{\prime}} G^{ss'A}_{\mathbf{k}^{\prime}}\left\langle U^{ss'}_{\mathbf{k}\mathbf{k}^{\prime}} U^{ss'}_{\mathbf{k}^{\prime}\mathbf{k}}\right\rangle_\mathrm{{imp}}\tilde{\mathbf{v}}_{\mathbf{k}^{\prime}}^{ss'},
\label{Eq_vel_si}
\end{equation}
Here $G^{ss'R}_{\mathbf{k}^{\prime}}$ and $G^{ss'A}_{\mathbf{k}^{\prime}}$ are retarded and advanced Green's functions respectively, given by 
\begin{align}
    G^{ss'R/A}_{\mathbf{k}} (\omega)= \frac{1}{\omega-\epsilon^{ss'}_\mathbf{k} \pm \frac{i\hbar}{2\tau_{ss'}}}
\end{align}
The ansatz $\tilde{\mathbf{v}}_{\mathbf{k}}^{ss'}= \eta^{ss'}\mathbf{v}_{\mathbf{k}}^{ss'}$ is substituted in Eq.~\ref{Eq_vel_si} to obtain the following solution for $\eta^{ss'}$:
\begin{align}
    &\eta_{ss'} = \frac{\mathcal{G}_{ss'}}{\mathcal{G}_{ss'}-\mathcal{H}_{ss'}},\end{align}
where     
\begin{align}
    \mathcal{G}_{ss'}=\int\limits_0^{2\pi} \frac{d\phi'}{2\pi}\langle U^{ss'}_{\mathbf{k}\mathbf{k}'} U^{ss'}_{\mathbf{k}'\mathbf{k}} \rangle_{\mathrm{imp}}\mathrm{, and }\hspace{1.5mm}
    \mathcal{H}_{ss'}\cos\phi=\int\limits_0^{2\pi} \frac{d\phi'}{2\pi}\cos\phi'\langle U^{ss'}_{\mathbf{k}\mathbf{k}'} U^{ss'}_{\mathbf{k}'\mathbf{k}} \rangle_{\mathrm{imp}}.
\end{align}
\section{Conductivity}
The quantum interference correction to conductivity is obtained by the calculation of a bare Hiakmi box and two dressed Hikami boxes.
The bare Hikami box at zero temperature is calculated as
\begin{equation}
     \sigma_{0}^{F}=\frac{e^{2}\hbar}{2\pi}\sum_{\textbf{q}}\boldsymbol{\Gamma}(\textbf{q})\sum_{\textbf{k}}\Tilde v_{\textbf{k}}^{ss'x}\Tilde v^{ss'x}_{\textbf{q-k}}G_{\mathbf{k}}^{ss'\mathrm{R}} G_{\mathbf{k}}^{ss'\mathrm{A}}G_{\mathbf{q-k}}^{ss'\mathrm{R}} G_{\mathbf{q-k}}^{ss'\mathrm{A}},
\end{equation}
In the small $\mathbf{q}$ limit, we find 
\begin{align}
    \sigma_0^F = -\frac{e^2 {s'}^2 \vartheta^2 N^{s'}_F \eta_{ss'}^2\tau_{ss'}^3}{\hbar^2} \sum_{\textbf{q}}\boldsymbol{\Gamma}(\textbf{q})
\end{align}
Here $\boldsymbol{\Gamma}(\textbf{q})$ is the vertex function which depends on $\mathbf{q}$ (incoming momentum) and must not be confused with the Gamma function $\Gamma(d)$. 

Two dressed Hikami boxes denoted as $\sigma_{R}^{F}$ and $\sigma_{A}^{F}$
\begin{align}
    \sigma_{R}^{F}= \frac{e^{2}\hbar}{2\pi}\sum_{\textbf{q}}\Gamma(\textbf{q})\sum_{\textbf{k}}\sum_{\textbf{k}_{1}}\Tilde v_{\textbf{k}}^{ssx}\Tilde v^{ssx}_{\textbf{q}-\textbf{k}_{1}}G_{\mathbf{k}}^{ss\mathrm{R}}G_{\mathbf{k_1}}^{ss\mathrm{R}}G_{\mathbf{q-k}}^{ss\mathrm{R}} G_{\mathbf{q-k_1}}^{ss\mathrm{R}}G_{\mathbf{k}}^{ss\mathrm{A}} G_{\mathbf{q-k_1}}^{ss\mathrm{A}}\langle U^{ss}_{\mathbf{k}_{1},\textbf{k}}U^{ss}_{\mathbf{q-k_{1}},\mathbf{q-k}}\bigl \rangle_{\mathrm{imp}},
\end{align}
\begin{align}
     \sigma_{A}^{F}= \frac{e^{2}\hbar}{2\pi}\sum_{\textbf{q}}\Gamma(\textbf{q})\sum_{\textbf{k}}\sum_{\textbf{k}_{1}}\Tilde v_{\textbf{k}}^{ssx}\Tilde v^{ssx}_{\textbf{q}-\textbf{k}_{1}}G_{\mathbf{k}}^{ss\mathrm{R}}G_{\mathbf{q-k_1}}^{ss\mathrm{R}}G_{\mathbf{k}}^{ss\mathrm{A}}G_{\mathbf{k_1}}^{ss\mathrm{A}}G_{\mathbf{q-k}}^{ss\mathrm{A}} G_{\mathbf{q-k_1}}^{ss\mathrm{A}}\langle U^{ss}_{\textbf{k},\mathbf{k}_{1}}U^{ss}_{\mathbf{q-k},\mathbf{q-k_1}}\bigl \rangle_{\mathrm{imp}},
\end{align}
We evaluate 
\begin{align}
    \sigma_{A}^{F}=\sigma_{R}^{F}=\frac{e^2 N^{s'}_F \tau_{ss'}^3\eta_{ss'}^2\vartheta^2{s'}^2 }{4\hbar^2 \mathcal{G}_{ss'}} (\textcolor{black}{\mathcal{A}^{ss'}_{\alpha-1}+\mathcal{A}^{ss'}_{\alpha+1}})  \sum_{\textbf{q}}\boldsymbol{\Gamma}(\textbf{q})
\end{align}
The ratio of dressed to bare Hikami box is given by 
\begin{align}
    \frac{\sigma_{A}^{F}}{\sigma_{0}^{F}} = -\frac{\textcolor{black}{\mathcal{A}^{ss'}_{\alpha-1}+\mathcal{A}^{ss'}_{\alpha+1}}}{4\mathcal{G}_{ss'}}.
\end{align}
The total conductivity is given by the sum of the bare and two dressed Hikami boxes: 
\begin{align}
    \sigma^F = -\frac{e^2 N^{s'}_F \tau_{ss'}^3\eta_{ss}^2\vartheta^2{s'}^2 }{4\hbar^2 \mathcal{G}_{ss'}} \left(1-\left(\frac{\textcolor{black}{\mathcal{A}^{ss'}_{\alpha-1}+\mathcal{A}^{ss'}_{\alpha+1}} }{2\mathcal{G}_{ss'}}\right)\right) \sum_{\textbf{q}}\boldsymbol{\Gamma}(\textbf{q})
\end{align}

\section{Bethe-Salpeter equation}
The Bethe-Salpeter equation for the vertex is given by
\begin{align}
    \boldsymbol{\Gamma}^{ss'}_{\mathbf{k}_{1}, \mathbf{k}_{2}}= \boldsymbol{\Gamma}_{\mathbf{k}_{1}, \mathbf{k}_{2}}^{ss'0}+\sum_{\mathbf{k}}\boldsymbol{\Gamma}_{\mathbf{k}_{1}, \mathbf{k}}^{ss'0} G_{\mathbf{k}}^{ss'R}G_{\mathbf{q}-\mathbf{k}}^{ss'A} \boldsymbol{\Gamma}^{ss'}_{\mathbf{k}, \mathbf{k}_{2}},
\end{align}
where the bare vertex $
\boldsymbol{\Gamma}^{ss'0}_{\mathbf{k_{1},k_{2}}}= \langle U^{ss'}_{\mathbf{k_1}\mathbf{k}_{2}}U^{ss'}_{\mathbf{-k_1}\mathbf{k_2}}\bigl\rangle_{\mathrm{imp}}$ takes the following form 
\begin{align}
\boldsymbol{\Gamma}^{ss'0}_{\mathbf{k_{1},k_{2}}} = \left(\frac{\hbar}{2\pi N^{s'}_F \mathcal{G}_{ss'} \tau_{ss'}}\right) \sum\limits_{m=0}^{4s} {\mathcal{A}^{ss'}_m} e^{im(\phi_1-\phi_2)},
\label{SI_Eq_Gamma0}
\end{align}
We assume the following ansatz for the vertex:
\begin{align}
\boldsymbol{\Gamma}^{ss'}_{\mathbf{k_{1},k_{2}}} = \left(\frac{\hbar}{2\pi N^{s'}_F \mathcal{G}_{ss'} \tau_{ss'}}\right) \sum\limits_{m=0}^{4s}\sum\limits_{n=0}^{4s} {\mathcal{V}^{ss'}_{mn}} e^{i(m\phi_1-n\phi_2)},
\label{SI_Eq_V_ansatz}
\end{align}
which solves the Bethe-Salpeter equation. The coefficients of the matrix $\mathcal{V}^{ss'}$ are given by the solution of the following equation: 
\begin{align}{\mathcal{V}}^{ss'} = (1-\mathcal{A}^{ss'} \Phi^{ss'}\mathcal{G}_{ss'}^{-1})^{-1} \mathcal{A}^{ss'},
\label{SI_Eq_V_matrix}
\end{align}
where
\begin{align}
    \Phi^{ss'}_{mn} = \int{\frac{d\phi}{2\pi} \frac{e^{i(n-m)\phi}}{1+i\tau_{ss'} \vartheta s' q \cos\phi}}
    = \left(1-\frac{Q^2}{2}\right) \delta_{mn} -\frac{iQ}{2}\left(\delta_{m,n+1}+\delta_{m,n-1}\right) 
    -\frac{Q^2}{4}\left(\delta_{m,n+2}+\delta_{m,n-2}\right),
\end{align}
and $Q=\vartheta\tau_{ss'} s'q$. It is possible to express the diagonal elements of the matrix $\mathcal{V}^{ss'}$ as 
\begin{align}
\mathcal{V}^{ss'}_{ii} = \frac{\mathcal{U}^{ss'}_{ii}}{\mathcal{W}^{ss'}_{ii}},
\end{align}
where 
\begin{align}
\mathcal{U}^{ss'}_{ii} &= \mathcal{G}_{ss'}\nonumber\\
\mathcal{W}^{ss'}_{ii} &=\left(-1+\frac{\mathcal{G}_{ss'}}{\mathcal{A}^{ss'}_{i}}\right)+\bigg( {2} \sum_j \frac{\mathcal{A}^{ss'}_j}{\mathcal{G}_{ss'}}
+\sum_{j<k} \alpha^{(2)}_{ss'jk} \frac{\mathcal{A}^{ss'}_j}{\mathcal{G}_{ss'}}\frac{\mathcal{A}^{ss'}_k}{\mathcal{G}_{ss'}}\nonumber\\
&\sum_{j<k<l}\alpha^{(3)}_{ss'jkl}\frac{\mathcal{A}^{ss'}_j}{\mathcal{G}_{ss'}}\frac{\mathcal{A}^{ss'}_k}{\mathcal{G}_{ss'}}\frac{\mathcal{A}^{ss'}_l}{\mathcal{G}_{ss'}} +  \sum_{j<k<l<m}\alpha^{(4)}_{ss'jklm}\frac{\mathcal{A}^{ss'}_j}{\mathcal{G}_{ss'}}\frac{\mathcal{A}^{ss'}_k}{\mathcal{G}_{ss'}}\frac{\mathcal{A}^{ss'}_l}{\mathcal{G}_{ss'}}\frac{\mathcal{A}^{ss'}_m}{\mathcal{G}_{ss'}}+...
+ \beta_{ss'}{\prod_j\frac{\mathcal{A}^{ss'}_j}{\mathcal{G}_{ss'}}}\bigg)\frac{Q^2}{D_{ss'}},
\end{align}
where
\begin{align}
D_{ss'} &= \prod_i\left(-1+\frac{\mathcal{G}_{ss'}}{\mathcal{A}^{ss'}_i}\right),
\end{align}
and the coefficients $\alpha$ and $\beta$ can be determined for specific cases. It is of interest to find the Cooperon gaps $(g^{ss'}_\alpha\equiv 2(-1+\mathcal{G}_{ss'}/\mathcal{A}^{ss'}_\alpha))$. Vanishing Cooperon gaps result in diverging elements $\mathcal{V}^{ss'}_{\alpha\alpha}$ in the limit $q\rightarrow 0$.

\section{The case $s=s'$}
Our focus here is the topmost conduction band with energy dispersion $\epsilon_\mathbf{k} = \hbar \vartheta s k$. In this case, it is possible to analytically find out the various coefficients introduced earlier. 
\begin{align}
&f^{ss}_m=\left(\frac{\Gamma(2s+1)}{\Gamma(m+1)\Gamma(2s+1-m)}\right)^{1/2}\nonumber\\
    &\mathcal{N}_{ss}= \frac{1}{\sqrt{2^{2s}}}\nonumber\\
    &\mathcal{F}^{ss}(\phi)= \cos^{4s}(\phi/2)\nonumber\\
    &\mathcal{G}_{ss}= \frac{\Gamma(2s+1/2)}{\sqrt{\pi}\Gamma(2s+1)}\nonumber\\
    &\mathcal{H}_{ss}= \frac{\Gamma(2s+1/2)}{\sqrt{\pi}(\Gamma(2s)+2s\Gamma(2s))}\nonumber\\
    &\eta^{ss} = 2s+1\nonumber\\
    &\mathcal{A}^{ss}_{0\leq m\leq 2s}= \frac{\Gamma(2s+1/2)}{\sqrt{\pi} \left(\prod\limits_{k=0}^{k=2s-m-1}{4s-m-k}\right) m!}\nonumber\\
    &\mathcal{A}^{ss}_{2s\leq m\leq 4s} =\mathcal{A}^{ss}_{4s-m},
\end{align}
where $\Gamma(x)$ is the Gamma-function. Remarkably, the wavefunction coefficients $f_m^{ss}$ are related to the square root of the entries of Pascal's triangle as shown in Fig.~\ref{fig:wfcoeff} of the main text.

Furthermore, the following condition guarantees that the Cooperon gap vanishes:
\begin{align}
g_\alpha^{ss}=\frac{\mathcal{G}_{ss}}{\mathcal{A}^{ss}_\alpha}=1.
\end{align}
The above condition is satisfied for $\alpha=2s$. Therefore in the limit $q\rightarrow 0$, the vertex correction is dominated by the following term:
\begin{align}
    \boldsymbol{\Gamma}^{ss'}_{\mathbf{k}_1\mathbf{k}_2} \sim  \frac{1}{q^2}e^{2is(\phi_1-\phi_2)}.
\end{align}
When $\phi_1-\phi_2\approx \pi$, the vertex carries a positive (negative) sign for integer (half-integer) values of $s$.  This implies weak localization for integer $s$ and weak antilocalization for half-integer $s$.

Furthermore, we recover the known results for graphene:
\begin{align}
&\eta^{\frac{1}{2}\frac{1}{2}}=2\nonumber\\
&\mathcal{F}^{\frac{1}{2}\frac{1}{2}}(\phi) = \cos^2(\phi/2)\nonumber\\
&\frac{\sigma_{A}^{F}}{\sigma_{0}^{F}} = -\frac{\textcolor{black}{\mathcal{A}^{\frac{1}{2}\frac{1}{2}}_{0}+\mathcal{A}^{\frac{1}{2}\frac{1}{2}}_{2}}}{4\mathcal{G}_{\frac{1}{2}\frac{1}{2}}} = -\frac{1}{4}.
\end{align}

\section{The case $s'\neq s$ and coefficient tables}
When $s'\neq s$, finding generalized analytical expressions is a cumbersome task. We explicitly evaluate the coefficients for the first few cases ($s\leq 7/2$). Table~\ref{tab:gsspO},~\ref{tab:gsspE},~\ref{tab:hsspO},~\ref{tab:hsspE} and present the values of the coefficients $\mathcal{G}^{ss'}$ and $\mathcal{H}^{ss'}$, respectively. The velocity correction coefficients are presented in Table~\ref{tab:etasspO} and ~\ref{tab:etasspE}. Interestingly, we find that the flat bands in the integer $s$ case have $\eta^{ss'}=0$, implying zero velocity correction. The bare Cooperon coefficients $A^{ss'}$ are presented in Table~\ref{tab:barecoopcoeffodd} and Table~\ref{tab:barecoopcoeffeven} for the half-integer $s$ and integer $s$ cases, respectively. We find that the value  $\alpha$ for which the Cooperon gap $g^{ss'}_\alpha$ vanishes is independent of $s'$. Therefore, when multiple bands cross the Fermi energy (for $s\geq 3/2$), each band results in the same qualitative behavior, i.e., localization for integer $s$ and antilocalization for half-integer $s$. 

\begin{table}[ht]
\centering
\begin{tabular}[t]{lcccccccc}
\toprule
${s'}\rightarrow$&$-\frac{7}{2}$&$-\frac{5}{2}$&$-\frac{3}{2}$&$-\frac{1}{2}$&$\frac{1}{2}$&$\frac{3}{2}$&$\frac{5}{2}$&$\frac{7}{2}$\\
$s\downarrow$ &\\
\toprule
$\frac{1}{2}$&--&--&--&$\frac{1}{2}$&$\frac{1}{2}$&--&--&--\\ \\
$\frac{3}{2}$&--&--&$\frac{5}{16}$&$\frac{5}{16}$&$\frac{5}{16}$&$\frac{5}{16}$&--&--\\ \\
$\frac{5}{2}$&--&$\frac{63}{256}$&$\frac{55}{256}$&$\frac{15}{64}$&$\frac{15}{64}$&$\frac{55}{256}$&$\frac{63}{256}$&--\\ \\
$\frac{7}{2}$&$\frac{429}{2048}$&$\frac{357}{2048}$&$\frac{349}{2048}$&$\frac{389}{2048}$&$\frac{389}{2048}$&$\frac{349}{2048}$&$\frac{357}{2048}$&$\frac{429}{2048}$\\ \\
\bottomrule
\end{tabular}
\caption{$\mathcal{G}^{ss'}$ for half-integer  pseudospin-$s$ fermions. }
\label{tab:gsspO}
\end{table}
\begin{table}[ht]
\centering
\begin{tabular}[t]{lccccccc}
\toprule
${s'}\rightarrow$&$-3$&$-2$&$-1$&$0$&$1$&$2$&$3$\\
$s\downarrow $\\
\toprule
$1$&--&--&$\frac{3}{8}$&$\frac{1}{2}$&$\frac{3}{8}$&--&--\\ \\
$2$&--&$\frac{35}{128}$&$\frac{1}{4}$&$\frac{11}{32}$&$\frac{1}{4}$&$\frac{35}{128}$&--\\ \\
$3$&$\frac{231}{1024}$&$\frac{49}{256}$&$\frac{199}{1024}$&$\frac{17}{64}$&$\frac{199}{1024}$&$\frac{49}{256}$&$\frac{231}{1024}$\\ \\ 
\bottomrule
\end{tabular}
\caption{$\mathcal{G}^{ss'}$ for integer pseudospin-$s$ fermions. }
\label{tab:gsspE}
\end{table}

\begin{table}[ht]
\centering
\begin{tabular}[t]{lcccccccc}
\toprule
${s'}\rightarrow$&$-\frac{7}{2}$&$-\frac{5}{2}$&$-\frac{3}{2}$&$-\frac{1}{2}$&$\frac{1}{2}$&$\frac{3}{2}$&$\frac{5}{2}$&$\frac{7}{2}$\\
$s\downarrow$ &\\
\toprule
$\frac{1}{2}$&--&--&--&$\frac{1}{4}$&$\frac{1}{4}$&--&--&--\\ \\
$\frac{3}{2}$&--&--&$\frac{15}{64}$&$\frac{7}{64}$&$\frac{7}{64}$&$\frac{15}{64}$&--&--\\ \\
$\frac{5}{2}$&--&$\frac{105}{512}$&$\frac{65}{512}$&$\frac{9}{128}$&$\frac{9}{128}$&$\frac{65}{512}$&$\frac{105}{512}$&--\\ \\
$\frac{7}{2}$&$\frac{3003}{16384}$&$\frac{1995}{16384}$&$\frac{1443}{16384}$&$\frac{851}{16384}$&$\frac{851}{16384}$&$\frac{1443}{16384}$&$\frac{1995}{16384}$&$\frac{3003}{16384}$\\ \\
\bottomrule
\end{tabular}
\caption{$\mathcal{H}^{ss'}$ for half-integer pseudospin-$s$ fermions. }
\label{tab:hsspO}
\end{table}
\begin{table}[ht]
\centering
\begin{tabular}[t]{lccccccc}
\toprule
${s'}\rightarrow$&$-3$&$-2$&$-1$&$0$&$1$&$2$&$3$\\
$s\downarrow $\\
\toprule
$1$&--&--&$\frac{1}{4}$&${0}$&$\frac{1}{4}$&--&--\\ \\
$2$&--&$\frac{7}{8}$&$\frac{1}{8}$&$0$&$\frac{1}{8}$&$\frac{7}{8}$&--\\ \\
$3$&$\frac{99}{512}$&$\frac{1}{8}$&$\frac{43}{512}$&$0$&$\frac{43}{512}$&$\frac{1}{8}$&$\frac{99}{512}$\\ \\ 
\bottomrule
\end{tabular}
\caption{$\mathcal{H}^{ss'}$ for integer pseudospin-$s$ fermions. }
\label{tab:hsspE}
\end{table}

\begin{table}[ht]
\centering
\begin{tabular}[t]{lcccccccc}
\toprule
${s'}\rightarrow$&$-\frac{7}{2}$&$-\frac{5}{2}$&$-\frac{3}{2}$&$-\frac{1}{2}$&$\frac{1}{2}$&$\frac{3}{2}$&$\frac{5}{2}$&$\frac{7}{2}$\\
$s\downarrow$ &\\
\toprule
$\frac{1}{2}$&--&--&--&$2$&$2$&--&--&--\\ \\
$\frac{3}{2}$&--&--&$4$&$\frac{20}{13}$&$\frac{20}{13}$&$4$&--&--\\ \\
$\frac{5}{2}$&--&$6$&$\frac{22}{9}$&$\frac{10}{7}$&$\frac{10}{7}$&$\frac{22}{9}$&$8$&--\\ \\
$\frac{7}{2}$&$8$&$\frac{136}{41}$&$\frac{2792}{1349}$&$\frac{3112}{2261}$&$\frac{3112}{2261}$&$\frac{2792}{1349}$&$\frac{136}{41}$&$8$\\ \\
\bottomrule
\end{tabular}
\caption{The velocity correction $\mathcal{\eta}^{ss'}$ for half-integer pseudospin-$s$ fermions. }
\label{tab:etasspO}
\end{table}
\begin{table}[ht]
\centering
\begin{tabular}[t]{lccccccc}
\toprule
${s'}\rightarrow$&$-3$&$-2$&$-1$&$0$&$1$&$2$&$3$\\
$s\downarrow $\\
\toprule
$1$&--&--&$3$&${1}$&$3$&--&--\\ \\
$2$&--&$5$&$2$&$1$&$2$&$5$&--\\ \\
$3$&$7$&$\frac{49}{17}$&$\frac{199}{113}$&$1$&$\frac{199}{113}$&$\frac{49}{17}$&$7$\\ \\ 
\bottomrule
\end{tabular}
\caption{The velocity correction $\mathcal{\eta}^{ss'}$ for integer pseudospin-$s$ fermions. }
\label{tab:etasspE}
\end{table}

\begin{table}[ht]
    \centering
    \begin{tabular}{lc|ccccccccccccc}
    \toprule
    &&$\mathcal{A}_0^{ss'}$ & $\mathcal{A}_1^{ss'}$ & $\mathcal{A}_2^{ss'}$ & $\mathcal{A}_3^{ss'}$ & $\mathcal{A}_4^{ss'}$ & $\mathcal{A}_5^{ss'}$ & $\mathcal{A}_6^{ss'}$ & $\mathcal{A}_7^{ss'}$ & $\mathcal{A}_8$& $\mathcal{A}_9^{ss'}$ & $\mathcal{A}_{10}^{ss'}$& &\\
    \midrule
    $s$& $s'$& \\
    \midrule
    $\frac{1}{2}$& $\frac{1}{2}$  & $\frac{1}{4}$&\color{blue}{$\frac{1}{2}$}&$\frac{1}{4}$\\ \\ 
    $\frac{1}{2}$& $-\frac{1}{2}$ & $\frac{1}{4}$&\color{blue}{$\frac{1}{2}$}&$\frac{1}{4}$    & & \\ \\
    $\frac{3}{2}$& $\frac{3}{2}$  & $\frac{1}{64}$&$\frac{3}{32}$&$\frac{15}{64}$ & \color{blue}{$\frac{5}{16}$}&$\frac{15}{64}$&$\frac{3}{32}$& $\frac{1}{64}$\\ \\ 
    $\frac{3}{2}$& $\frac{1}{2}$ & $\frac{9}{64}$&$\frac{3}{32}$&$\frac{7}{64}$ &  \color{blue}{$\frac{5}{16}$}&$\frac{15}{64}$&$\frac{3}{32}$& $\frac{1}{64}$    & & \\ \\
    $\frac{3}{2}$& $-\frac{1}{2}$  & $\frac{9}{64}$&$\frac{3}{32}$&$\frac{7}{64}$ &  \color{blue}{$\frac{5}{16}$}&$\frac{7}{64}$&$\frac{3}{32}$& $\frac{9}{64}$\\ \\ 
    $\frac{3}{2}$& $-\frac{3}{2}$ & $\frac{1}{64}$&$\frac{3}{32}$&$\frac{15}{64}$ & \color{blue}{$\frac{5}{16}$}&$\frac{15}{64}$&$\frac{3}{32}$& $\frac{1}{64}$\\ \\ 
     $\frac{5}{2}$& $\frac{5}{2}$ & $\frac{1}{1024}$&$\frac{5}{512}$&$\frac{45}{1024}$ & $\frac{15}{128}$&$\frac{105}{512}$&\color{blue}{$\frac{63}{256}$}& $\frac{105}{512}$    &$\frac{15}{128}$ & $\frac{45}{1024}$ & $\frac{5}{512}$&$\frac{1}{1024}$\\ \\
     $\frac{5}{2}$& $\frac{3}{2}$ & $\frac{25}{1024}$&$\frac{45}{512}$&$\frac{101}{1024}$ & $\frac{7}{128}$&$\frac{65}{512}$&\color{blue}{$\frac{55}{256}$}& $\frac{65}{512}$    &$\frac{7}{128}$ & $\frac{101}{1024}$ & $\frac{45}{512}$&$\frac{25}{1024}$\\ \\
     $\frac{5}{2}$& $\frac{1}{2}$ & $\frac{25}{256}$&$\frac{5}{128}$&$\frac{21}{256}$ & $\frac{3}{32}$&$\frac{9}{128}$&\color{blue}{$\frac{15}{64}$}& $\frac{9}{128}$    &$\frac{3}{32}$ & $\frac{21}{256}$ & $\frac{5}{128}$&$\frac{25}{256}$\\ \\
     $\frac{5}{2}$& $-\frac{1}{2}$ & $\frac{25}{256}$&$\frac{5}{128}$&$\frac{21}{256}$ & $\frac{3}{32}$&$\frac{9}{128}$&\color{blue}{$\frac{15}{64}$}& $\frac{9}{128}$    &$\frac{3}{32}$ & $\frac{21}{256}$ & $\frac{5}{128}$&$\frac{25}{256}$\\ \\
     $\frac{5}{2}$& $-\frac{3}{2}$ & $\frac{25}{1024}$&$\frac{45}{512}$&$\frac{101}{1024}$ & $\frac{7}{128}$&$\frac{65}{512}$&\color{blue}{$\frac{55}{256}$}& $\frac{65}{512}$    &$\frac{7}{128}$ & $\frac{101}{1024}$ & $\frac{45}{512}$&$\frac{25}{1024}$\\ \\
     $\frac{5}{2}$& -$\frac{5}{2}$ & $\frac{1}{1024}$&$\frac{5}{512}$&$\frac{45}{1024}$ & $\frac{15}{128}$&$\frac{105}{512}$&\color{blue}{$\frac{63}{256}$}& $\frac{105}{512}$    &$\frac{15}{128}$ & $\frac{45}{1024}$ & $\frac{5}{512}$&$\frac{1}{1024}$\\ \\
    \bottomrule
    \end{tabular}
    \caption{Bare Cooperon coefficients $\mathcal{A}^{ss'}$ for half-integer pseudospin-$s$ fermions. The highlighted blue color indicates $\mathcal{A}^{ss'}_\alpha=\mathcal{G}_{ss'}$, which is the condition for vanishing Cooperon gap $g_\alpha$.}
    \label{tab:barecoopcoeffodd}
\end{table}

\begin{table}[ht]
    \centering
    \begin{tabular}{lc|ccccccccccccc}
    \toprule
    &&$\mathcal{A}_0^{ss'}$ & $\mathcal{A}_1^{ss'}$ & $\mathcal{A}_2^{ss'}$ & $\mathcal{A}_3^{ss'}$ & $\mathcal{A}_4^{ss'}$ & $\mathcal{A}_5^{ss'}$ & $\mathcal{A}_6^{ss'}$ & $\mathcal{A}_7^{ss'}$ & $\mathcal{A}_8^{ss'}$& $\mathcal{A}_9^{ss'}$ & $\mathcal{A}_{10}^{ss'}$& $\mathcal{A}_{11}^{ss'}$ & $\mathcal{A}_{12}^{ss'}$\\
    \midrule
    $s$& $s'$& \\
    \midrule
    ${1}$& $1$  & $\frac{1}{16}$&{$\frac{1}{4}$}&\color{blue}{$\frac{3}{8}$}&$\frac{1}{4}$&$\frac{1}{16}$\\ \\ 
    ${1}$& $0$  & $\frac{1}{4}$&0&\color{blue}{$\frac{1}{2}$}&0&$\frac{1}{4}$\\ \\ 
    ${1}$& $-1$  & $\frac{1}{16}$&{$\frac{1}{4}$}&\color{blue}{$\frac{3}{8}$}&$\frac{1}{4}$&$\frac{1}{16}$\\ \\
    ${2}$& $2$  & $\frac{1}{256}$&{$\frac{1}{32}$}&$\frac{7}{64}$&$\frac{7}{32}$&\color{blue}{$\frac{35}{128}$}&$\frac{7}{32}$&$\frac{7}{64}$&{$\frac{1}{32}$}& $\frac{1}{256}$\\ \\ 
    ${2}$& $1$  & $\frac{1}{16}$&{$\frac{1}{8}$}&$\frac{1}{16}$&$\frac{1}{8}$&\color{blue}{$\frac{1}{4}$}&$\frac{1}{8}$&$\frac{1}{16}$&{$\frac{1}{8}$}& $\frac{1}{16}$\\ \\
    ${2}$& $0$  & $\frac{9}{64}$&0&{$\frac{3}{16}$}&0& \color{blue}{$\frac{11}{32}$}&0&$\frac{3}{16}$&0&$\frac{9}{64}$\\ \\
    ${2}$& $-1$  & $\frac{1}{16}$&{$\frac{1}{8}$}&$\frac{1}{16}$&$\frac{1}{8}$&\color{blue}{$\frac{1}{4}$}&$\frac{1}{8}$&$\frac{1}{16}$&{$\frac{1}{8}$}& $\frac{1}{16}$\\ \\
    ${2}$& $-2$  & $\frac{1}{256}$&{$\frac{1}{32}$}&$\frac{7}{64}$&$\frac{7}{32}$&\color{blue}{$\frac{35}{128}$}&$\frac{7}{32}$&$\frac{7}{64}$&{$\frac{1}{32}$}& $\frac{1}{256}$\\ \\
    ${3}$& $3$ &$\frac{1}{4096}$& $\frac{3}{1024}$& $\frac{33}{2048}$&$\frac{55}{1024}$&$\frac{495}{4096}$&$\frac{99}{512}$&\color{blue}{$\frac{231}{1024}$}&$\frac{99}{512}$&$\frac{495}{4096}$&$\frac{55}{1024}$&$\frac{33}{2048}$&$\frac{3}{1024}$&$\frac{1}{4096}$\\ \\
    ${3}$& $2$ &$\frac{9}{1024}$& $\frac{3}{64}$& $\frac{47}{512}$&$\frac{5}{64}$&$\frac{55}{1024}$&$\frac{1}{8}$&\color{blue}{$\frac{49}{256}$}&$\frac{1}{8}$&$\frac{55}{1024}$&$\frac{5}{64}$&$\frac{47}{512}$&$\frac{3}{64}$&$\frac{9}{1024}$\\ \\
    ${3}$& $1$ &$\frac{225}{4096}$& $\frac{75}{1024}$& $\frac{65}{2048}$&$\frac{95}{1024}$&$\frac{271}{4096}$&$\frac{43}{512}$&\color{blue}{$\frac{199}{1024}$}&$\frac{43}{512}$&$\frac{271}{4096}$&$\frac{95}{1024}$&$\frac{65}{2048}$&$\frac{75}{1024}$&$\frac{225}{4096}$\\ \\
     ${3}$& $0$ &$\frac{25}{256}$& $0$& $\frac{15}{128}$ &$0$ &$\frac{39}{256}$ &$0$ &\color{blue}{$\frac{17}{64}$} &$0$&$\frac{39}{256}$&$0$&$\frac{15}{128}$&$0$&$\frac{25}{256}$\\ \\
    ${3}$& $-1$ &$\frac{225}{4096}$& $\frac{75}{1024}$& $\frac{65}{2048}$&$\frac{95}{1024}$&$\frac{271}{4096}$&$\frac{43}{512}$&\color{blue}{$\frac{199}{1024}$}&$\frac{43}{512}$&$\frac{271}{4096}$&$\frac{95}{1024}$&$\frac{65}{2048}$&$\frac{75}{1024}$&$\frac{225}{4096}$\\ \\
    ${3}$& $-2$ &$\frac{9}{1024}$& $\frac{3}{64}$& $\frac{47}{512}$&$\frac{5}{64}$&$\frac{55}{1024}$&$\frac{1}{8}$&\color{blue}{$\frac{49}{256}$}&$\frac{1}{8}$&$\frac{55}{1024}$&$\frac{5}{64}$&$\frac{47}{512}$&$\frac{3}{64}$&$\frac{9}{1024}$\\ \\
    ${3}$& $-3$ &$\frac{1}{4096}$& $\frac{3}{1024}$& $\frac{33}{2048}$&$\frac{55}{1024}$&$\frac{495}{4096}$&$\frac{99}{512}$&\color{blue}{$\frac{231}{1024}$}&$\frac{99}{512}$&$\frac{495}{4096}$&$\frac{55}{1024}$&$\frac{33}{2048}$&$\frac{3}{1024}$&$\frac{1}{4096}$\\ \\
    \bottomrule
    \end{tabular}
    \caption{Bare Cooperon coefficients $\mathcal{A}^{ss'}$ for integer pseudospin-$s$ fermions. The highlighted blue color indicates $\mathcal{A}^{ss'}_\alpha=\mathcal{G}_{ss'}$, which is the condition for vanishing Cooperon gap $g_\alpha$.}
    \label{tab:barecoopcoeffeven}
\end{table}

\begin{table}[ht]
    \centering
    \begin{tabular}{lc|cc}
    \toprule
    &&$\alpha$& $\textcolor{black}{4/X^{ss'}_\alpha}$\\
    \midrule
    $s$& $s'$& \\
    \midrule
    ${1}/{2}$& ${1}/{2}$  & 1& 1\\ 
    ${1}/{2}$& $-{1}/{2}$  & 1& 1\\ 
    ${1}$& ${1}$  & 2&1/2\\ 
    ${1}$& ${0}$  & 2&1\\ 
    ${1}$& ${-1}$  & 2&1/2\\ 
    ${3}/{2}$& ${3}/{2}$  & 3&5/16\\ 
    ${3}/{2}$& ${1}/{2}$  & 3&13/16\\ 
    ${3}/{2}$& $-{1}/{2}$  & 3&13/16\\ 
    ${3}/{2}$& $-{3}/{2}$  & 3&5/16\\ 
    ${2}$& ${2}$  & 4& 7/32\\ 
    ${2}$& ${1}$  & 4& 1/2\\
    ${2}$& ${0}$  & 4& 5/8\\ 
    ${2}$& ${-1}$  & 4&1/2\\ 
    ${2}$& ${-2}$  & 4& 7/32\\ 
    ${5}/{2}$& ${5}/{2}$  & 5&21/128\\ 
    ${5}/{2}$& ${3}/{2}$  & 5&45/128\\ 
    ${5}/{2}$& ${1}/{2}$  & 5&21/32\\ 
    ${5}/{2}$& $-{1}/{2}$  & 5&21/32\\ 
    ${5}/{2}$& $-{3}/{2}$  & 5&45/128\\ 
    ${5}/{2}$& $-{5}/{2}$  & 5&21/128\\ 
    ${3}$& ${3}$  & 6&33/256\\ 
    ${3}$& ${2}$  & 6&17/64\\ 
    ${3}$& ${1}$  & 6&113/256\\ 
    ${3}$& ${0}$  & 6&29/64\\ 
    ${3}$& ${-1}$  & 6&113/256\\ 
    ${3}$& ${-2}$  & 6&17/64\\ 
    ${3}$& ${-3}$  & 6&33/256\\ 
    \bottomrule
    \end{tabular}
    \caption{The values  $\alpha$ such that $g_\alpha^{ss'}=0$, and the corresponding $X^{ss'}_\alpha$}
    \label{tab:xalpha}
\end{table}

\clearpage
\section{Magnetoconductivity}
The zero-field conductivity from $|\mathbf{k}ss'\rangle$ is finally evaluated to be 
\begin{align}
    \sigma_{ss'} = -\sum_\alpha\frac{e^2}{\pi h} Y^{ss'}_\alpha \int {d(q^2) \frac{1}{l_{ss'\alpha}^{-2}+q^2}},
    \end{align}
    where 
    \begin{align}
    & Y^{ss'}_\alpha = \frac{{\eta^{ss'}}^{2} s'^{2}}{4 X^{ss'}_\alpha {\mathcal{G}^{ss'}}^2} \left(1-\frac{\mathcal{A}_1^{ss'}}{2\mathcal{G}^{ss'}}\right),\nonumber\\
    & l_{ss'\alpha}^{-2} = \frac{g^{ss'}_\alpha}{2X^{ss'}_\alpha {l_{ss'}}^2},\nonumber\\
    & {l_{ss'}}^{2-} = \frac{2}{\vartheta^2 \tau_{ss'}^2}
\end{align}
and the vertex $\boldsymbol{\Gamma}_{\mathbf{q}}^{ss'}$ is expressed as 
\begin{align}
\boldsymbol{\Gamma_\mathbf{q}}^{ss'} = \frac{\hbar}{2\pi N_F^{s'}\tau_{ss'}\mathcal{G}^{ss'}}\sum_\alpha\frac{2}{{g^{ss'}_\alpha}^2 + X^{ss'}_\alpha Q^2} e^{i\alpha\pi}
\end{align}
The vanishing Cooperon gaps will result in the most dominant contribution to the conductivity. The values of $\alpha$ such that $g_\alpha^{ss'}=0$, and the corresponding $X^{ss'}_\alpha$ are presented in Table~\ref{tab:xalpha}. In the weak-field limit, the magnetoconductivity is given by 
\begin{align}
    \Delta \sigma(B)_{ss'} = \frac{e^2}{\pi h} \sum_\alpha Y^{ss'}_\alpha \bigg[ \Psi\left(\frac{l_B^2}{l_\phi^2} + \frac{l_B^2}{l_{ss'\alpha}^2}+\frac{1}{2} \right)- \log\left(\frac{l_B^2}{l_\phi^2} + \frac{l_B^2}{l_{ss'\alpha}^2}\right)\bigg],
\end{align}
where $\Psi(x)$ is the digamma function, $l_\phi$ is the coherence length, and $l_B = \sqrt{\hbar/4eB}$ is the magnetic length of a Cooperon. 
\end{widetext}

\bibliography{biblio.bib}
\end{document}